\documentclass[journal, ,onecolumn,12pt]{IEEEtran}
\usepackage{amsmath,amsfonts}
\usepackage{array}
\usepackage[caption=false,font=normalsize,labelfont=sf,textfont=sf]{subfig}
\usepackage{textcomp}
\usepackage{stfloats}
\usepackage{url}
\usepackage{verbatim}
\usepackage{graphicx}
\usepackage{cite}
\usepackage{booktabs}
\usepackage{url}            
\usepackage{amsfonts}       
\usepackage{xcolor}         
\usepackage{wrapfig}
\usepackage{graphicx}
\usepackage{amsmath}
\usepackage[ruled]{algorithm2e}
\usepackage{bbm}
\usepackage{stmaryrd,scalerel}
\usepackage[makeroom]{cancel}
\DeclareMathOperator*{\argmin}{argmin}
\usepackage{amssymb}

\makeatletter
\DeclareMathAlphabet\mathbfcal{OMS}{cmsy}{b}{n} 

\usepackage[flushleft]{threeparttable}

\begin{document}
\title{Scalable Multivariate Fronthaul Quantization for Cell-Free Massive MIMO}

\author{Sangwoo Park,~\IEEEmembership{Member,~IEEE,}  Ahmet Hasim Gokceoglu, Li Wang,~\IEEEmembership{Senior Member,~IEEE}, and 
        Osvaldo Simeone,~\IEEEmembership{Fellow,~IEEE}

\thanks{S. Park and O. Simeone are with the King’s Communications, Learning \& Information Processing (KCLIP) lab within the Centre for Intelligent Information Processing Systems (CIIPS), Department of Engineering, King’s College London, London WC2R 2LS, U.K. (e-mail: \{sangwoo.park, osvaldo.simeone\}@kcl.ac.uk). A. H. Gokceoglu and L. Wang are with Huawei’s Sweden Research and Development Center, Stockholm, Sweden (e-mail: ahmet.hasim.gokceoglu1@huawei.com, powerking@live.co.uk).} 
\thanks{The work of O. Simeone was supported by European Union’s Horizon Europe project CENTRIC (101096379), by the Open Fellowships of the EPSRC (EP/W024101/1), and by the EPSRC project (EP/X011852/1).} 
}

\maketitle
\begin{abstract}
The conventional approach to the fronthaul design for cell-free massive MIMO system follows the compress-and-precode (CP) paradigm. Accordingly, encoded bits and precoding coefficients are shared by the  distributed unit (DU) on the fronthaul links, and precoding takes place at the radio units (RUs). Previous theoretical work has shown that CP can be potentially improved by a significant margin by \emph{precode-and-compress} (PC) methods, in which all baseband processing is carried out at the DU, which compresses the precoded signals for transmission on the fronthaul links. The theoretical performance gain of PC methods are particularly pronounced when the DU implements multivariate quantization (MQ), applying joint quantization across the signals for all the RUs.  However, existing solutions for MQ are characterized by a computational complexity that grows exponentially with the sum-fronthaul capacity from the DU to all RUs. This work sets out to design scalable MQ strategies for PC-based cell-free massive MIMO systems. For the low-fronthaul capacity regime, we present $\alpha$-parallel MQ ($\alpha$-PMQ), whose complexity is exponential only in the fronthaul capacity towards an individual RU, while performing close to full MQ. $\alpha$-PMQ tailors MQ to the topology of the network by allowing for parallel  local quantization steps for RUs that do not interfere too much with each other. For the high-fronthaul capacity regime, we then introduce neural MQ, which replaces the  exhaustive search in MQ with gradient-based updates for  a neural-network-based decoder, attaining a complexity  that grows linearly with the sum-fronthaul capacity. Numerical results  demonstrate that the proposed scalable MQ strategies outperform CP for both the low and high-fronthaul capacity regimes at the cost of increased computational complexity at the DU (but not at the RUs).
\end{abstract}
\section{Introduction}
\label{sec:intro}

\subsection{Context and Motivation}
\begin{figure}[t]
    \begin{center}\includegraphics[scale=0.26]{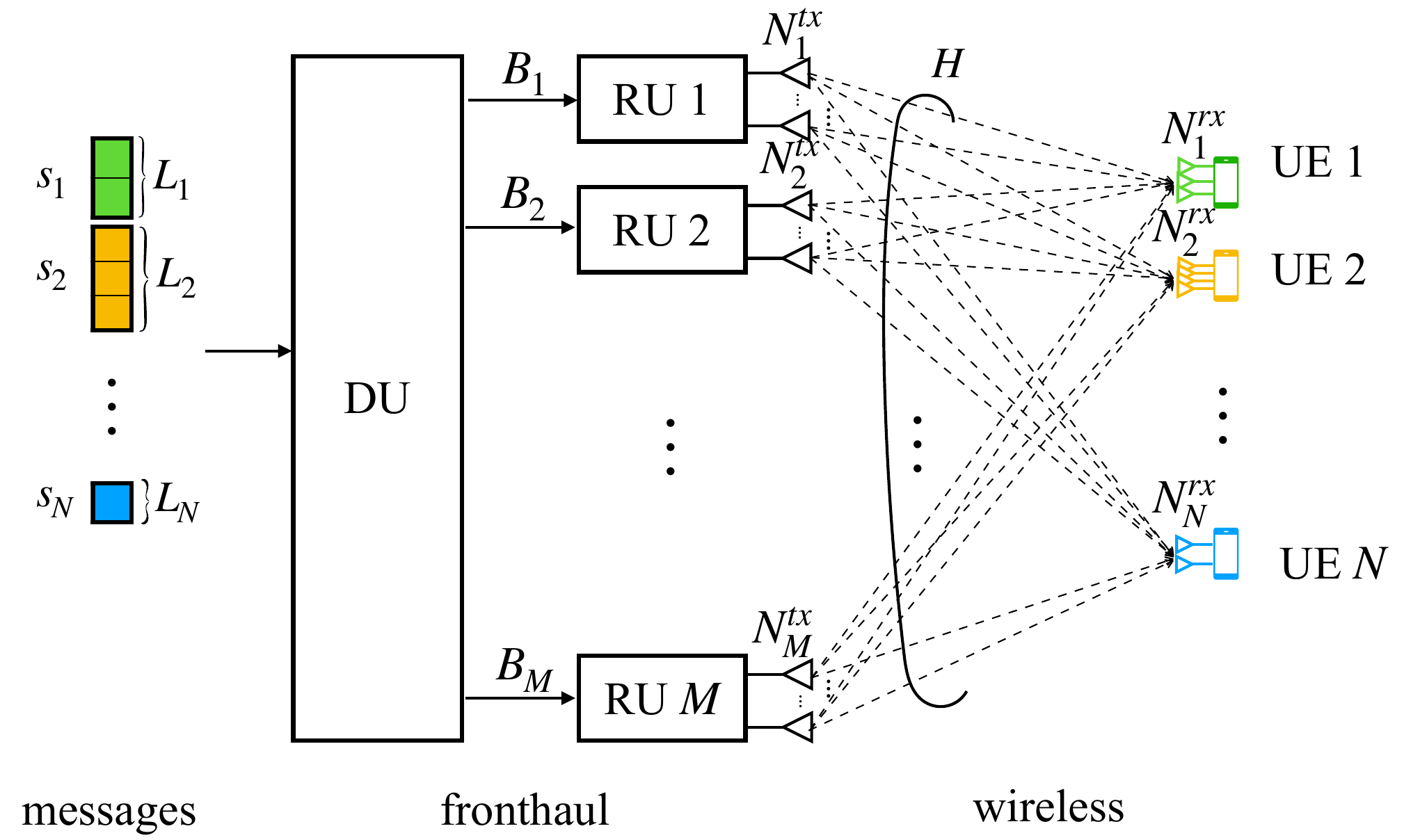}
      \end{center}
      \caption{Cell-free massive MIMO architecture considered in this project, consisting of $N$ multi-antenna UEs, $M$ multi-antenna RUs, and a DU. All baseband processing is done at the DU, which carries out compression of the baseband signals for transmission over capacity-limited fronthaul links.}
      \label{fig:system_model_general}
   \end{figure}

In modern wireless systems, base stations are disaggregated into radio units (RUs), distributed units (DUs), and central units (CUs), with RUs and DUs connected via fronthaul links. As shown in Fig.~\ref{fig:system_model_general}, a  DU may control several RUs, supporting coordinated transmission and reception across multiple distributed RUs \cite{simeone2012cooperative, ngo2024ultra}. While several functional splits between DUs and RUs have been defined, current deployments adopt a specific split, typically referred to as 7.2x, whereby lower-physical layer (PHY) functionalities such as precoding are carried out closer to the antennas, at the RU, while higher-PHY functionalities such as encoding are implemented at the DU \cite{rodriguez2020cloud, grönland2023learningbased}. This functional split becomes problematic for regimes characterized by massive antenna arrays and large spectral efficiencies \cite{rodriguez2020cloud}.  This issue is currently one of the key factors limiting the deployment of cell-free massive MIMO  systems based on disaggregated base stations, such as O-RAN \cite{5g_oran_spectrum, ngo2024ultra}.
 
\begin{figure*}[t]
    \begin{center}\includegraphics[scale=0.26]{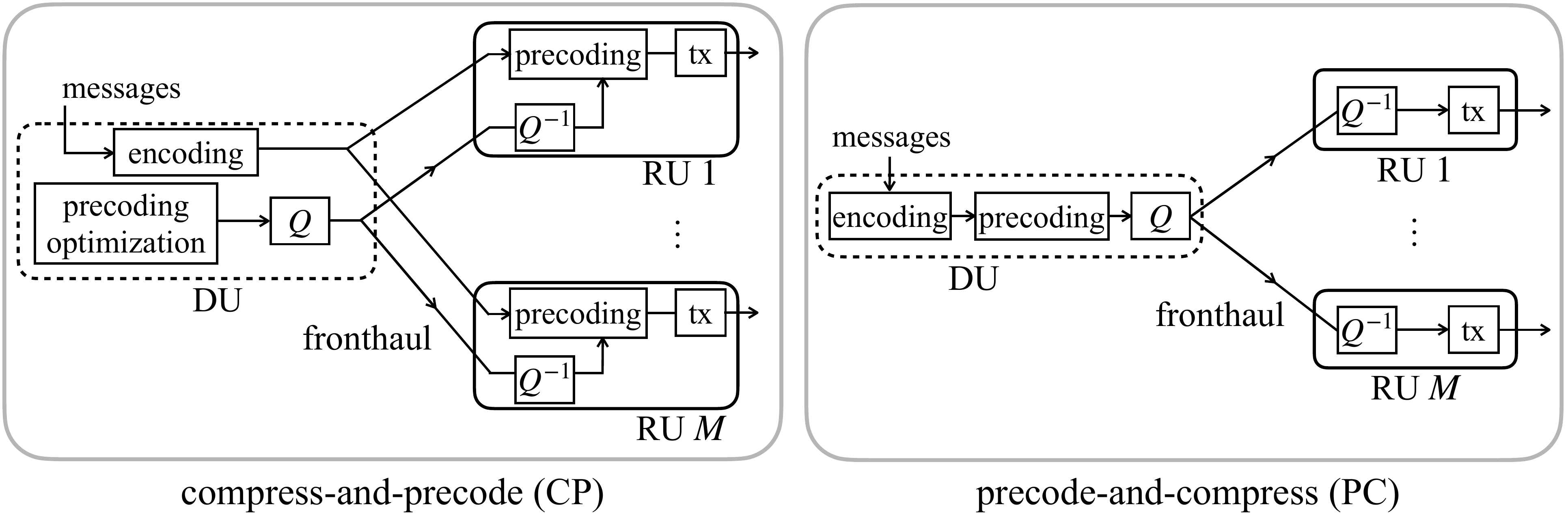}
      \end{center}
      \caption{(Left) In the conventional compress-and-precode (CP) scheme, the DU transmits information bits to all RUs, as well as the corresponding compressed precoding matrix to each RU. (Right) In the precode-and-compress (PC) scheme, the DU transmits the respective precoded and compressed baseband vector to each RU. This paper studies novel multivariate compression strategies for PC.}
      \label{fig:comp_split}
   \end{figure*}

In the downlink, the 7.2x functional split prescribes an approach that may be referred to as \emph{compress-and-precode} (CP). As shown in Fig.~\ref{fig:comp_split} (left), with CP, the DU  applies channel coding to all the information bits, and evaluates   the precoding matrices, which are transmitted in a compressed form to every RU via the corresponding  fronthaul link. Hence,  the fronthaul overhead of this approach may be substantial due to the separate transmission of precoding matrices and coded bits.

To mitigate this problem, an alternative functional split has been introduced  in which the DU applies coding and precoding,  transmitting to each RU the respective complex baseband signals after compression \cite{park2013joint, lee2016multivariate, qiao2024meta}. This way, as suggested by the theoretical results in \cite{park2013joint,sanderovich2009uplink, simeone2012cooperative}, the fronthaul capacity requirements may be drastically reduced.  We refer to such an approach as  \emph{precode-and-compress} (PC), which is the subject of this paper and is illustrated in Fig.~\ref{fig:comp_split} (right). 

While the PC functional split has the potential to effectively lessen the fronthaul capacity, the theoretical results in \cite{park2013joint} require complex baseband compression schemes at the DU, whose practical implementation is an open problem. Some steps in this direction were reported in \cite{lee2016multivariate}, which proposes \emph{multivariate quantization} (MQ), whereby the  baseband  signals for all RUs are quantized jointly \cite{lee2016multivariate}.

Specifically, reference  \cite{lee2016multivariate} proposed a data-driven algorithm for MQ that was shown to outperform existing per-RU point-to-point quantization techniques. However, the algorithm in \cite{lee2016multivariate} suffers from a computational complexity that grows  exponentially in the fronthaul sum-rate, and  is also limited to the case in which all the UEs and RUs are equipped with single antennas. The goal of this paper is to design practical, scalable PC-based solutions for distributed large-scale MIMO systems.

\subsection{Related Work}
{\color{black} Recent work on fronthaul design for cell-free massive MIMO has aimed at  (\emph{i}) enhancing CP; (\emph{ii}) improving PC; and (\emph{iii}) proposing hybrid methods between  CP and PC, corresponding  different functional splits.  

Representative papers on CP include \cite{khorsandmanesh2022quantization},  which addresses precoder design  by taking into account the limited fronthaul capacity; as well as \cite{demir2024cell, bashar2020exploiting}, which consider optimizing the allocation of resources. Recent advances in PC have focused on linear dimension reduction techniques \cite{arad2018precode, wiffen2021distributed, qiao2024meta} followed by uniform quantization. These schemes require  sharing the linear transformation matrix, which entails additional fronthaul communication overhead, which may become substantial in cell-free massive MIMO systems. This limitation has been recently alleviated via meta-learning \cite{qiao2024meta}. 

Lastly, alternatives to CP and PC have been studied  \cite{liu2015graph, kang2015fronthaul, murti2022learning} that showcase the optimality of different functional splits depending  on the underlying dynamics of the system. These schemes require the solution of  non-convex optimization problems or lower-complexity methods such as reinforcement learning \cite{murti2022learning}.

As a final remark, the scalability of cell-free massive MIMO systems has been theoretically demonstrated in  \cite{bjornson2020scalable, parida2022cell} by allowing for dynamic matching of DUs and UEs.
}

Our work complements studies on precoding designs such as \cite{khorsandmanesh2022quantization,  arad2018precode, wiffen2021distributed, qiao2024meta} by focusing solely on fronthaul compression. It contributes to the line of work on scalable cell-free massive MIMO by showing for the first time the practical feasibility of MQ-based PC functional splits.


\subsection{Main Contribution}
{\color{black} This work introduces two new MQ techniques for PC-based transmission in  cell-free massive MIMO systems. The main contributions of this work can be summarized as follows:
\begin{itemize}
    \item We introduce \emph{$\alpha$-parallel multivariate quantization} ($\alpha$-PMQ), a novel MQ scheme that  has computational complexity growing exponentially only in the per-RU fronthaul rate, while demonstrating a small performance gap with respect to MQ \cite{lee2016multivariate} in the low-fronthaul capacity regime. $\alpha$-PMQ tailors MQ to the topology of the network by allowing for parallel {\color{black} local quantization steps for RUs that do not interfere too much with each other.}
    \item Furthermore, we introduce \emph{neural-multivariate quantization} (neural-MQ), another novel MQ scheme with computational complexity that grows linearly in the fronthaul sum-rate. Neural-MQ replaces the  exhaustive search in MQ with gradient-based updates for  a neural-network-based decoder. We demonstrate that neural-MQ outperforms  CP, as well as an infinite precoding benchmark that assumes linear precoding,  in the high-fronthaul capacity regime. 
\end{itemize}
}

The rest of the paper is organized as follows. In Sec.~\ref{sec:sys_model}, we describe the general system model of the cell-free MIMO systems, and review the state-of-the-art PC-based solution \cite{lee2016multivariate} designed for single-antenna UEs and RUs in Sec.~\ref{sec:wonju}. We then propose $\alpha$-PMQ in Sec.~\ref{sec:PMQ} and neural-MQ in Sec.~\ref{sec:neural_MQ}.  Experimental results are provided in Sec.~\ref{sec:experiments}, and Sec.~\ref{sec:conclusion} concludes the paper.


\section{System Model}
\label{sec:sys_model}
\subsection{Setting}
As shown in Fig.~\ref{fig:system_model_general}, in the cell-free setting of interest, a set $\mathcal{M}=\{1,...,M\}$ of RUs communicates to $N$ UEs through $M$ RUs. The $m$-th RU has ${N^\text{tx}_{m}}$ {transmit antennas}, and is connected to the DU through a {fronthaul link with capacity} ${B_m}$ bits per channel use, i.e., the fronthaul link carries $B_m$ bit/s/Hz when normalized by the bandwidth of the radio interface. Each $n$-th UE has $N_n^\text{rx}$ antennas. We denote the overall number of transmit antennas as $N^{\text{tx}}$, i.e., $N^\text{tx} = \sum_{m=1}^M N^\text{tx}_m$, and the overall number of receive antennas as $N^\text{rx}=\sum_{n=1}^N N^\text{rx}_n$.


The {DU} wishes to transmit an $L_n\times 1$ vector $s_n^k \in \mathbb{C}^{L_n}$ of $L_n$ complex symbols to each $n$-th UE for $n=1,...,N$ on each channel use $k$. The information symbols are assumed to be zero mean and independent, and we normalize their powers such that the equality $\mathbb{E}[s_n^k (s_n^k)^\dagger]=I_n$ holds for all $n=1,...,N$, where $I_n$ is the $L_n \times L_n$ identity matrix. 

Denote as $H_n^k$ the $N_n^\text{rx}\times N^\text{tx}$ channel matrix between all RUs and UE $n$ for channel use $k$, and as 
\begin{align}
    H^k = \big[(H_1^k)^\top,...,(H_N^k)^\top\big]^\top    
\end{align}
the overall channel matrix of size $ N^\text{rx} \times N^\text{tx}$. The $N^\text{tx}\times L$ precoding matrix 
\begin{align}
    W^k=[W_1^k,...,W_N^k]    
\end{align}
collects all the $N^\text{tx}\times L_n$ precoding matrices $W_n^k$ used to precode symbols $s_n^k$ towards user $n$ for $n=1,...,N$ for channel use $k$. The {precoding matrices} $\{W^k \} $  are optimized by the DU based on knowledge of the channel matrices $\{H^k\}$. 


Given the precoding matrix $W_k$ for channel use $k$, and further denoting $W_{m,n}^k$ as the $N^\text{tx}_m \times L_n$ precoding matrix used by RU $m$ to communicate with UE $n$ which gives the overall precoding matrix $W_n^k = [(W_{1,n}^k)^\top, ..., (W_{M,n}^k)^\top]^\top$ for user $n$, the $N^\text{tx} \times 1$ precoded symbol vector is given by
\begin{align}  \label{eq:x=Ws}
    {x}^k= \sum_{n=1}^N W_n^k s_n^k = W^k{s^k}
\end{align}
as a function of the symbol vector ${s^k}=[(s_1^k)^\top,...,(s_N^k)^\top]^{\top}$.  We partition the precoded vector $x^k$ in (\ref{eq:x=Ws}) across the RUs as
\begin{align} \label{eq:partitioned_x_k}
    x^k = \begin{bmatrix}
            x_1^k  \\
            \vdots \\
            x_M^k
            \end{bmatrix},
\end{align} 
where 
\begin{align}
    x_m^k = W_m^k s^k,
\end{align}
denoting as $W_{m}^k=[W_{m,1}^k, ..., W_{m,N}^k]$ the precoding matrix associated to RU $m$. Note the slight abuse of notation, as $W_n^k$ represents the beamforming matrix for UE $n$ and $W_m^k$ the beamforming matrix for RU $m$. 

Due to fronthaul capacity limitations, as we will detail in the rest of the section, the $N_m^\text{tx} \times 1$  symbol vector transmitted by RU $m$ at channel use $k$, denoted as $\hat{x}_m^k$, may differ from the precoded vector $x_m^k$ in (\ref{eq:partitioned_x_k}). We impose the average power constraint
\begin{align} \label{eq:avg_power_constraint}
    \mathbb{E}[ || \hat{x}_m^k ||^2 ] \leq P_m
\end{align}
for each RU $m$, where the expectation is taken with respect to transmit symbols $s^k$. Given the transmit vector $\hat{x}_m^k$ for all $m=1,...,M$, the $N_n^\text{rx} \times 1$ received signal vector $y^k_n$ at the UE $n$ is given by
\begin{align}
    \label{eq:rx_basic}
    y^k_n &=  H^k_{n} \hat{x} + z^k_n,
\end{align}
where $z^k_n \sim \mathcal{CN}(0, \sigma^2 I)$ is the $N_n^\text{rx} \times 1$ complex Gaussian noise vector. 

Given an $N_n^\text{rx}\times L_n$ receive beamforming matrix $F_n^k$, UE $n$ estimates the transmitted signal at channel use $k$ as \begin{align} \label{eq:symbol_estimation}
    \hat{s}_n^k=(F_n^k)^\dagger y^k_n,
\end{align}
for all $n=1,...,N$.  The receive beamforming matrix $F_n^k$ is generally designed by UE $n$ based on the available channel state information. We will address this point in Sec.~\ref{subsec:EI_with_CE}.

\subsection{{\color{black} Compress-and-Precode}}
In the conventional  \emph{compress-and-precode} (CP) strategy, the DU applies an entry-wise uniform quantizer $Q^\text{CP}(\cdot)$ to the precoding matrix $W_{m}^k$ for each RU $m$, producing the quantized precoding matrix 
\begin{align} \label{eq:compressed_W}
\hat{W}_{m}^k = Q^\text{CP}(W_{m}^k).    
\end{align}
The quantizer has a resolution of $B^\text{CP}$ bits per entry.  The DU then sends the bits describing the quantized matrix $W_{m}^k$ to RU $m$ on the fronthaul along with the information bits for all $N$ UEs to the $m$-th RU on the fronthaul. Each $m$-th RU then transmits the signal 
\begin{align} \label{eq:CP_tx}
    \tilde{x}_m^k = \gamma_m \hat{W}^k_{m}s^k,
\end{align}
where $\gamma_m$ is a parameter introduced to satisfy the power constraint (\ref{eq:avg_power_constraint}).


{\color{black} While  precoding matrix is in principle designed per-channel-use basis,} in order to reduce the fronthaul overhead, CP typically shares the same precoding matrix across multiple $K^\text{CP}$ channel uses\cite{lorca2013lossless, grönland2023learningbased}. 
Increasing $K^\text{CP}$ generally entails a trade-off between quality of precoding, which increases with a smaller $K^\text{CP}$, and fronthaul overhead, which decreases as $K^\text{CP}$ grows larger. In fact,  the fronthaul capacity constraint imposes the inequality 
\begin{align} \label{eq:CP_constraint}
    B_m \geq   \underbrace{R_\text{sum}}_{\substack{\text{sum-rate}\\\text{(bit/channel use)}}}  \cdot \underbrace{\frac{1}{R_\text{code}}}_{\text{code rate}} +  \underbrace{N_m^\text{tx}\cdot L\cdot B^\text{CP}}_{\text{precoding quantization}} \cdot \underbrace{\frac{1}{K^\text{CP}}}_{\substack{\text{precoding}\\\text{reuse factor}}},
\end{align}
where $R_\text{sum}$ is the sum-rate, in  bits per channel use,  across all $N$ UEs, and  $R_\text{code} \leq 1$ is the channel coding rate. The first term, $R_\text{sum}/R_\text{code}$, in (\ref{eq:CP_constraint}) accounts for the transmission of all coded bits to each RU, while the second term accounts for precoding information.


\subsection{{\color{black} Precode-and-Compress}} \label{sec:PC}
Unlike CP strategies, PC methods \cite{park2013joint, lee2016multivariate} compress directly the precoded vector \eqref{eq:x=Ws}. As a result, the quantized vector is given by 
\begin{align} \label{eq:hat_x}
    \hat{{x}}^k=Q({x}^k)=[(\hat{x}_1^k)^\top,...,(\hat{x}_M^k)^\top]^\top   
\end{align}
for some quantization function $Q(\cdot)$ with resolution $B_m$ bits. Note that function $Q(\cdot)$ may apply jointly across all entries of the vector \cite{lee2016multivariate}. We denote as $b_m^k \in \{0,1\}^{B_m}$  the discrete index identifying the quantized signal $\hat{x}_m^k$ transmitted by DU to RU $m$ on the fronthaul link at channel use $k$.

The mapping between bits $b_m^k$ and baseband symbols $\hat{x}_m^k$, which is to be applied by RU $m$, is defined by an inverse quantization function $f_m: \{0,1\}^{B_m} \rightarrow \mathbb{C}^{N_m^\text{tx}}$
\begin{align}
    \hat{x}_m^k = f_m(b_m^k).
\end{align} We will further denote as $f(b_{1:M}^k)=[ f_1(b_1^k)^\top, ..., f_M(b_M^k)^\top ]^\top$ the collection of the quantized outputs for all RUs, i.e., $\hat{x}^k= f(b_{1:M}^k)$ with $b_{1:M} = \{b_m\}_{m=1}^M$. 

After recovering the dequantized vector $\hat{x}_m^k$ from the bits $b_m^k$, the RU $m$ transmits the $N^\text{tx}_{m}\times 1$ vector 
\begin{align} \label{eq:hat_x_m}
\tilde{x}_m^k=\gamma_m \hat{x}_m^k = \gamma_m [\hat{x}_{m,1}^k,...,\hat{x}^k_{m,N^\text{tx}_{m}}]^\top,    
\end{align}
in which we denote as $\hat{x}^k_{m,i}$ the quantized precoded symbol transmitted at the $i$-th antenna of RU $m$ at channel use $k$, and the parameter $\gamma_m$ ensures the power constraint (\ref{eq:avg_power_constraint}) as  for CP in (\ref{eq:CP_tx}). Combining the signals transmitted by all RUs, we can write (\ref{eq:hat_x_m}) as 
\begin{align} \label{eq:Gamma_all_tx}
    \tilde{x}^k = \Gamma \hat{x}^k,
\end{align}
where $\Gamma = \text{diag}(\gamma_1 \mathbf{1}_1, ..., \gamma_M \mathbf{1}_M)$ with $\mathbf{1}_m$ being the all-one vector of size $N_m^\text{tx}$.

\subsection{{\color{black} Precode-and-Compress} with Conventional Quantization} \label{sec:VQ}
In this subsection, we introduce a benchmark transmission scheme based on PC {\color{black} method} and conventional \emph{vector quantization} (VQ)  \cite{lee2016multivariate}. Throughout the rest of the paper,  we omit the channel use index $k$  to simplify the notation. Conventional VQ is applied in a point-to-point manner,  separately for each RU,  yielding a quantized vector $\hat{x}_m$ for each precoded signal $x_m$ for RU $m$. 

To this end, let us fix a quantization codebook $\hat{\mathcal{X}}_m = \{\hat{x}_{m,j}\}_{j=1}^{2^{B_m}}$ containing $2^{B_m}$ vectors $\hat{x}_{m,j}$ of size $N_m^\text{tx} \times 1$, as well as an arbitrary mapping $f(b_m|\hat{\mathcal{X}}_m)$ between $B_m$ bits $b_m$ and codewords in set $\hat{\mathcal{X}}_m$. VQ finds the quantized vector $\hat{x}_m=f(b_m|\hat{\mathcal{X}}_m)$ for each RU $m$ as
\begin{align} \label{eq:VQ}   
    b_{m}^\text{VQ} = \argmin_{ b_m \in \{0,1\}^{B_m} } ||  x_{m} - f(b_m|\hat{\mathcal{X}}_m) ||^2,
\end{align} 
and we denote as
\begin{align}
    \hat{x}_m &= Q^\text{VQ}(x_m|\hat{\mathcal{X}}_m) = f(b_m^\text{VQ}|\hat{\mathcal{X}}_m).
\end{align}


The $2^{B_m}$ vectors in the  codebook $\hat{\mathcal{X}}_m=\{\hat{x}_{m,j}\}_{j=1}^{2^{B_m}}$ can be ideally optimized for each RU $m$ by addressing the problem of minimizing the average quantization error, i.e.,  \cite{lee2016multivariate}
\begin{align}
    &\min_{\hat{\mathcal{X}}_m} \mathbb{E}\big[ || x_m -  Q^\text{VQ}(x_m|\hat{\mathcal{X}}_m) ||^2 \big],
\end{align}
where the average is taken over both symbols $s$ and channels $H$. 

In practice, the quantization codebook $\hat{\mathcal{X}}_m$ is shared across many coherence intervals, and is updated only when the channel statistics change significantly. Therefore, the codebook is modified only occasionally, and need not be accounted for when evaluating the fronthaul overhead of PC.

\section{State of the Art on Multivariate Quantization}\label{sec:wonju}
This section summarizes the state-of-the-art fronthaul \emph{multivariate quantization} (MQ) schemes introduced in \cite{lee2016multivariate} for PC. Reference \cite{lee2016multivariate} assumed single-antenna RUs, i.e., $N^\text{tx}_{m}=1$, single-antenna UEs, i.e., $N^\text{rx}_n=1$, and a single message per UE, i.e., $L_n=1$, and it focused on a single channel use. Accordingly, in this section, we set $N^\text{tx}_{m}=N^\text{rx}_n=L_n=1$ for all $n=1,...,N$, and we drop the channel use superscript $k$. Furthermore, since the receive beamforming matrix $F_n^k$ in  (\ref{eq:symbol_estimation}) reduces to a scalar gain, we set it without loss of generality to $F_n^k=1$. Finally, reference \cite{lee2016multivariate} assumed an average power constraint $\mathbb{E}[||\hat{x}_m^k||^2] \leq P_m$ over both channels $H$ and transmitted symbols $s^k$, which is less strict than the per-channel average power constraint (\ref{eq:avg_power_constraint}), and thus here we modify the MQ scheme in \cite{lee2016multivariate} to satisfy the constraint (\ref{eq:avg_power_constraint}).

\subsection{Multivariate Quantization} \label{subsec:MQ_given_codebook}
Given a fronthaul capacity $B_m$ for the $m$-th RU,  the quantization codebook $\hat{\mathcal{X}}_m$  contains $2^{B_m}$ quantization codewords, i.e., $\hat{\mathcal{X}}_m = \{\hat{x}_{m,j}\}_{j=1}^{2^{B_m}}$ (see Sec.~\ref{sec:VQ}), which represent the possible values of the  quantized signals \eqref{eq:hat_x_m} transmitted by RU $m$. Each codeword $\hat{x}_m$ is indexed by $B_m$ bits $b_m \in \{0,1\}^{B_m}$. We further  denote as $\hat{\mathcal{X}}_{1:M}=\{\hat{\mathcal{X}}_{m}\}_{m=1}^M$  the collection of codebooks for all $M$ RUs. MQ  \cite{lee2016multivariate} implements a quantization function $Q(\cdot)$ in (\ref{eq:hat_x}) that  selects the codeword index $b_m$ to  transmit through the fronthaul link to each RU $m$ via an exhaustive search over the $2^{\sum_{m=1}^M B_m }$ combination of codewords. 

Specifically,  MQ implements the quantization function 
\begin{align}
    \label{eq:obj_MQ_sota}
     b_{1:M}^\text{MQ} = \argmin_{ b_{1:M} \in \{0,1\}^{\sum_{m=1}^M B_m}} \sum_{n=1}^N \text{EI}_n ( x, f(b_{1:M}|\hat{\mathcal{X}}_{1:M}) ),
\end{align}
and we denote as
\begin{align}
    \hat{x} &= Q^\text{MQ}(x|\hat{\mathcal{X}}_{1:M}) = f(b_{1:M}^\text{MQ}|\hat{\mathcal{X}}_{1:M}) 
\end{align}
where $\text{EI}_n(x,\hat{x})$ is the effective interference (EI) caused by the choice of codewords $\hat{x}$ on UE $n$.  The EI criterion will be introduced in the next subsection.

\subsection{Effective Interference}
To complete the description of the MQ rule in (\ref{eq:obj_MQ_sota}), we now introduce the EI objective function \cite{lee2016multivariate}.  To this end, we write the average effective SINR for UE $n$ as
\begin{align}
    \label{eq:average_sinr_sota}
    \text{SINR}_n = \frac{ |H_n W_{n}|^2 }{\sigma^2 +  \underbrace{\Big|H_n \Big(\Gamma\hat{x}-x + \sum_{n' \neq n}W_{n'}s_{n'}\Big)\Big|^2}_{\substack{=\text{ effective interference, EI}}}},
\end{align}
where $\sigma^2$ stands for the variance of the complex Gaussian noise at the UEs and we have used (\ref{eq:Gamma_all_tx}). The numerator in \eqref{eq:average_sinr_sota} represents the average power of the useful part of the received signal, $H_n W_n s_n$, assuming no quantization,  while the denominator in the sum of the noise power, $\sigma^2$, and of the EI.


 The EI for UE $n$ is the power of the quantization error and of the interference affecting reception at UE $n$. From 
 \eqref{eq:average_sinr_sota},  the EI can be expressed as 
\begin{align}
\label{eq:EI_def}
\text{EI}_n (x,\hat{x}) = \big|H_n \big(\Gamma\hat{x}-W_{n}s_n\big)\big|^2. 
\end{align} 
The notation $\text{EI}_n (x, \hat{x})$ makes explicit the dependence of the EI for each UE $n$ on the quantized signal $\hat{x}$.

\subsection{Sequential Multivariate Quantization}
\label{sec:smq}
As mentioned, problem  \eqref{eq:obj_MQ_sota} requires an exhaustive search over a space that is exponentially large in the sum-fronthaul rate $\sum_{m=1}^M B_m$, i.e.,  the complexity is of the order $\mathcal{O}(2^{\sum_{m=1}^M B_m})$ (see Table~\ref{table:tab_comp_comp}). To address this  computational complexity issue,  inspired by the information-theoretic optimality of sequential encoding in the regime of infinitely long blocklengths, reference \cite{lee2016multivariate} proposed a \emph{sequential MQ} (SMQ) scheme.

With SMQ, the DU applies  \emph{local quantization} for the signals to be sent to each RU on the fronthaul by searching only over the respective codebooks of the RUs. This search is done sequentially, one RU at a time, by following an arbitrary order over the RUs. Furthermore, quantization for each RU takes  into account the outputs of the local quantization steps for the previously considered RUs in the given order. 

Mathematically, assuming for illustration the order  $m=1,2,...,M$,  over the RU index $m$, SMQ obtains the $m$-th component $\hat{x}_m$ of the quantized signal $\hat{x}$ as\begin{align}
    \label{eq:smq}
    b_{m}^\text{seq} = \argmin_{ b_m \in \{0,1\}^{B_m}  } \sum_{n=1}^N  \text{EI}_n (x, f( b^{(m)}_{1:M} |\hat{\mathcal{X}}_{1:M}) ), \end{align}
and we denote as 
\begin{align}
    \hat{x}_m &= Q^\text{seq}(x|\hat{\mathcal{X}}_{m}, \hat{x}_1,...,\hat{x}_{m-1}) = f(b_m^\text{seq} |\hat{\mathcal{X}}_m ), 
\end{align}
with $f(b_{1:M}^{(m)}|\hat{\mathcal{X}}_{1:M}) = [\hat{x}_1, ..., \hat{x}_{m-1},  f(b_m|\hat{\mathcal{X}}_m), 0, ..., 0]^\top$ containing the quantized signals for the previously considered RUs. Accordingly, the signal for RU $m$ is obtained by fixing the quantized signals $\{\tilde{x}_{m'} = \gamma_{m'} \hat{x}_{m'} \}_{m'=1}^{m-1}$ for the previously considered RUs, while setting the other transmitted signals to zero. After addressing problem (\ref{eq:smq}) sequentially for all RUs,  the quantized vector is obtained as $\hat{x} = [\hat{x}_1, ..., \hat{x}_M]^\top$ as in \eqref{eq:hat_x}.

The complexity of SMQ is linear in the number of RUs, scaling as $\mathcal{O}(\sum_{m=1}^M 2^{B_m})$, since each problem \eqref{eq:smq} requires a search over the $2^{B_m}$ codewords in codebook $\hat{\mathcal{X}}_m$ (see Table~\ref{table:tab_comp_comp}).



\begin{table}
    \renewcommand\arraystretch{1.4}
  \caption{{\color{black} Computational complexity of PC methods ($M$ is the total number of RUs, $B_m$ [bits/s/Hz] is the fronthaul capacity between DU and RU $m$;  $T$ is the number of iterations of $\alpha$-PMQ; $I$ is the number of GD steps of neural-MQ; $D_m$ and  $K_m$ are  the number of hidden neurons and hidden layers of neural-network-based codebook for RU $m$ employed by neural-MQ)}}
  \label{table:tab_comp_comp}
  \centering
  \begin{tabular}{l|l}
    {\color{black}PC scheme} & {\color{black} computational complexity} \\ 
    \hline
    VQ   & $\mathcal{O}( 2^{ B_m})$      \\
    MQ      & $\mathcal{O}( 2^{\sum_{m=1}^M B_m})$      \\
    SMQ         & $\sum_{m=1}^M \mathcal{O}( 2^{B_m})$ \\
    $\alpha$-PMQ        & $T\cdot \mathcal{O}( 2^{ \max_{m} B_m})  $  \\
    neural-MQ &  $2 I\cdot \mathcal{O}( \sum_{m=1}^M  D_m(B_m + D_m(K_m-1) + 2N_m^\text{tx} )$ 
  \end{tabular}
\end{table}

\subsection{Codebook Optimization} \label{sec:codebook_opt_conven}
In order to design the codebooks $\hat{\mathcal{X}}_{1:M}$, reference \cite{lee2016multivariate} used a {data-driven} algorithm, namely Lloyd-Max, also known as $K$-means \cite{simeone2022machine}, that alternates between solving $M$ convex problems and addressing the exponential-complexity MQ problem \eqref{eq:obj_MQ_sota}.

Mathematically, the design objective for the codebooks reads
\begin{align}
    \label{eq:obj_wonju_codebook}
    \min_{\hat{\mathcal{X}}_{1:M}} \mathbbm{E}\Bigg[ \sum_{n=1}^N \text{EI}_n(x, \hat{x}) \Bigg],
\end{align}
where $\hat{x}$ is obtained from either MQ (\ref{eq:obj_MQ_sota}) or SMQ (\ref{eq:smq}), and  the average is taken over both symbols $s$ and channels $H$.

Problem (\ref{eq:obj_wonju_codebook}) is addressed by optimizing over codebooks $\hat{\mathcal{X}}_{1:M}$ -- a convex problem -- and over the quantization step (\ref{eq:obj_MQ_sota}) in an alternating fashion.

\section{$\alpha$-Parallel Mutlivariate Quantization}\label{sec:PMQ}


As discussed in Sec.~\ref{sec:smq}, SMQ reduces the computational complexity of MQ via sequential local quantization steps at the RUs for a number of iterations equal to the number of RUs. However, this approach was shown in \cite{lee2016multivariate} to have a significant performance gap as compared to MQ.

\begin{figure*}
    \begin{center}\includegraphics[scale=0.12]{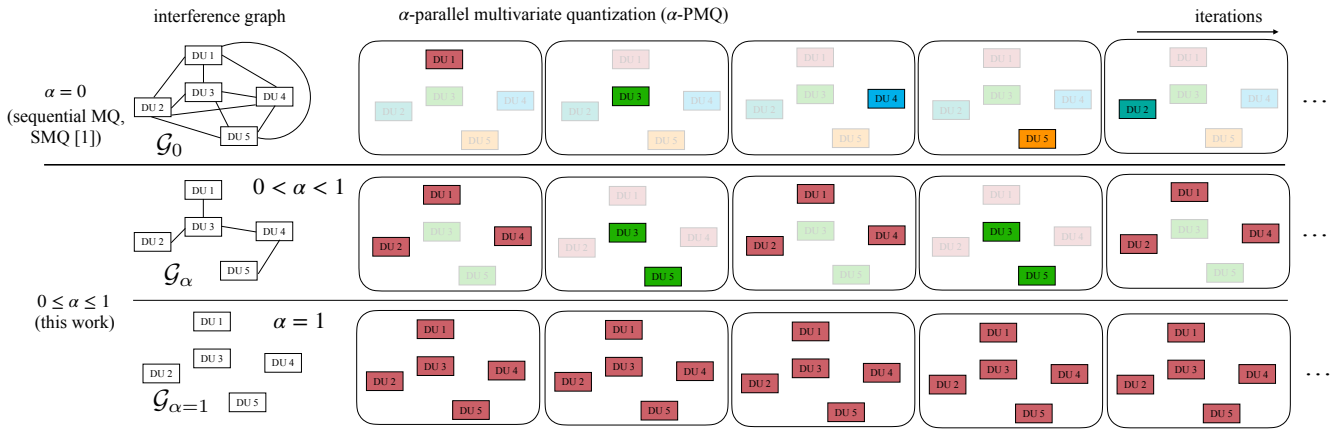}
      \end{center}
      \caption{Previous work on reduced complexity MQ introduced SMQ, which applies sequential local quantization (for a number of iterations equal to the number of RUs) \cite{lee2016multivariate}. The proposed $\alpha$-parallel multivariate quantization ($\alpha$-PMQ) scheme allows for the signals of multiple RUs to be updated using local quantization in parallel. To determine the local quantization schedule, depending on the hyperparameter $\alpha$, an interference graph with RUs as nodes is constructed. At each iteration (square box), for all RUs in an independent set of the interference graph, the DU  carries out local quantization in parallel. An edge between RUs is included in the interference graph if the level of interference  between RUs is larger then a threshold determined by $\alpha$.
      }\label{fig:alpha_MPMQ_illust}
   \end{figure*}

 
 In this section, we generalize SMQ by proposing \emph{$\alpha$-parallel multivariate quantization} ($\alpha$-PMQ), which shares with SMQ the complexity of order $\mathcal{O}(\sum_{m=1}^M 2^{B_m})$ for quantization (see Table~\ref{table:tab_comp_comp}), while significantly reducing the performance gap with respect to MQ. As illustrated in Fig.~\ref{fig:alpha_MPMQ_illust}, the main idea of $\alpha$-PMQ is to allow for parallel local quantization steps at RUs that do not interfere too much with each other. Accordingly,  $\alpha$-PMQ is tailored to the topology of the network. We continue to focus on the single antenna case in order to maintain consistency with \cite{lee2016multivariate}, while the extension to multiple antenna case will be discussed in Sec.~\ref{subsec:alpha_pmq_for_multi_antenna}. 
 
\subsection{EI Decomposition}
We start by rewriting the EI in \eqref{eq:EI_def} by isolating the separate contribution of all RUs. Specifically,  we write
\begin{align}
    \label{eq:EI_decomp}
    \text{EI}_n(x,\hat{x}) = \bigg|\sum_{m=1}^M  \underbrace{ H_{m,n} (\gamma_m \hat{x}_m - W_{m,n}s_n)  }_{=\Delta_{m \rightarrow n}}\bigg|^2, 
\end{align}
where the scalar $\Delta_{m \rightarrow n}$ represents the \emph{disturbance from RU $m$ to UE $n$}; $H_{m,n}$ is the channel between RU $m$ and UE $n$; and $W_{m,n}$ is the precoding used by RU $m$ to communicate to UE $n$. With definition \eqref{eq:EI_decomp}, the sum-EI criterion for MQ in \eqref{eq:obj_MQ_sota} can be equivalently rewritten as
\begin{align}
    \label{eq:sum_EI_with_Delta}
    \sum_{n=1}^N  \text{EI}_{n}(x,\hat{x}) = \bigg|\bigg| \sum_{m=1}^M \Delta_m \bigg|\bigg|^2,
\end{align}
where we have denoted $\Delta_m = [\Delta_{m\rightarrow 1}^\top, ..., \Delta_{m\rightarrow N}^\top]^\top$ for the $N\times 1$ \emph{disturbance vector from RU $m$} to all $N$ UEs.

\subsection{$\alpha$-Parallel Multivariate Quantization}
As illustrated in Fig.~\ref{fig:alpha_MPMQ_illust}, $\alpha$-PMQ operates sequentially like SMQ. However, unlike SMQ, at each iteration, rather than evaluating the quantized signal for a single RU, $\alpha$-PMQ applies parallel local quantization steps for a suitably chosen subset of RUs. In this regard, SMQ can be viewed as a special case of $\alpha$-PMQ in which only one RU applies local quantization at each iteration.

Let us define as $\mathcal{M}_{\alpha}^{(t)}$ the subset of RUs $\{1,...,M\}$ for which the DU updates the quantized signals at iteration $t$. How to construct the sequence of these subsets will be discussed in the next subsection. At each iteration $t=1,...,T$, for each RU $m \in \mathcal{M}_\alpha^{(t)}$, the DU chooses a quantized symbol $\hat{x}_{m}^{(t)} \in \hat{\mathcal{X}}_m$, while for all RUs $m \notin \mathcal{M}_{\alpha}^{(t)}$ it sets  $\hat{x}_m^{(t)}=\hat{x}_m^{(t-1)}$. All quantized signals are initialized to be equal to zero. Given the updated quantized signals, for all RUs $m\in \mathcal{M}_\alpha^{(t)}$,  the disturbance vector $\Delta_m^{(t)}=[(\Delta_{m \rightarrow 1}^{(t)})^\top,..., (\Delta_{m \rightarrow N}^{(t)})^\top]^\top$ is updated  accordingly using definition \eqref{eq:EI_decomp}, as \begin{equation}\label{eq:alpha_pmq_Delta_update}\Delta_{m \rightarrow n}^{(t)}= H_{m,n}(\gamma_m\hat{x}_m^{(t)}-W_{m,n}s_n).\end{equation} For all RUs $m \notin \mathcal{M}_{\alpha}^{(t)}$, the DU sets  $\Delta_{m \rightarrow n}^{(t)}=\Delta_{m \rightarrow n}^{(t-1)}$. 

We now discuss how to update the quantized signals at any iteration $t$. Given the disturbance vector $\Delta_{m}^{(t-1)}$ from the previous iteration, and subset $\mathcal{M}_{\alpha}^{(t)}$, $\alpha$-PMQ solves in parallel the problems 
\begin{align}
    \label{eq:local_quant}
    b_{m}^{\text{$\alpha$-PMQ},(t)} = \argmin_{ b_m \in \{0,1\}^{B_m}  } \big|\big|{\Delta}_{\neg m}^{(t-1)} + \Delta_m(  f(b_m|\hat{\mathcal{X}}_m)  )\big|\big|^2, \end{align}
and we denote as
\begin{align}
    \hat{x}_m^{(t)} &= Q^\text{$\alpha$-PMQ}(x|\hat{\mathcal{X}}_{m}, \Delta_1^{(t-1)}, ..., \Delta_M^{(t-1)}) = f(b_m^{\text{$\alpha$-PMQ},(t)} |\hat{\mathcal{X}}_m ) 
\end{align}
for all RUs $m \in \mathcal{M}_\alpha^{(t)}$, where we have defined as
\begin{align}
    {\Delta}_{\neg m}^{(t-1)}=\sum_{m' \neq m} \Delta_{m'}^{(t-1)}
\end{align}
the \emph{disturbance from RUs other than $m$}, along with 
\begin{align}
    \Delta_m(\hat{x}_m)= [(\Delta_{m \rightarrow 1}(\hat{x}_m))^\top,..., (\Delta_{m\rightarrow N}(\hat{x}_m))^\top ]^\top, 
\end{align}
where $\Delta_{m \rightarrow n}(\hat{x}_m) = H_{m,n}(\gamma_m \hat{x}_m-W_{m,n}s_n)$ as defined in \eqref{eq:EI_decomp}. Problem (\ref{eq:local_quant}) is a modified version of the original objective of MQ in (\ref{eq:obj_MQ_sota})  in which the contributions to the EI from all RUs other than $m$ are fixed to the values obtained at the previous iteration. One can directly check that  by setting $\mathcal{M}_\alpha^{(t)} = \{t\}$ and $T=M$, $\alpha$-PMQ recovers SMQ.

\subsection{Optimizing the Update Schedule}
The main idea behind the selection of subsets $\mathcal{M}_\alpha^{(t)}$ is to allow for parallel local quantization steps \eqref{eq:local_quant} only for RUs that do not interfere too much with each other. The intuition is that RUs whose transmitted signals affect significantly the  same subset of devices should be optimized jointly, which is facilitated by sequential, rather than parallel, optimization. 

To elaborate, let us define the $N\times 1$ channel gain vector for node $m$ as
\begin{align}
    \label{eq:interference_graph_1}
    G_m = \Big[ |H_{m,1}|^2, ..., |H_{m,N}|^2 \Big]^\top.
\end{align}
Then, we build an  \emph{interference graph} $\mathcal{G}_\alpha= (\mathcal{M}, \mathcal{E}_\alpha)$ by defining the edge set $\mathcal{E}_\alpha$ as 
\begin{align}
    \label{eq:interference_graph_2}
    (m, m') \in \mathcal{E}_\alpha \text{ if } G_m^T G_{m'} > g(\alpha) \text{ and } m \neq m',
\end{align}
with the threshold $g(\alpha)$  defined to ensure that the number of the edges $|\mathcal{E}_\alpha|$ is a fraction $1-\alpha$ of the total number of edges $\sum_{i=1}^{M-1} i$, i.e., $|\mathcal{E}_\alpha| = \lceil (1-\alpha)(\sum_{i=1}^{M-1}i) \rceil$, with ceiling operation $\lceil \cdot \rceil$. Mathematically, the threshold is defined as
\begin{align}
    \label{eq:interference_graph_3}
    g(\alpha) = {G_m^T G_{m'}}_{(\lfloor \alpha(\sum_{i=1}^{M-1} i ) \rfloor )}
\end{align}
denoting as ${G_m^\top G_{m'}}_{(k)}$  the $k$-th smallest value in the set $\{\{ G_m^\top G_{m'}\}_{m=1}^M\}_{m'=m+1}^M$ with ${G_m^\top G_{m'}}_{(0)}=-1$.  For $\alpha=1$ the interference graph is disconnected, while for $\alpha=0$, it becomes fully connected.

Given the interference graph, $\alpha$-PMQ seeks for subsets of nodes that are not connected by edges, indicating that the corresponding RUs do not interference to much with each other. To this end, the algorithm obtains \emph{independent sets} $\mathcal{M}_\alpha \subseteq \mathcal{M}$. By definition, an independent set $\mathcal{M}_\alpha \subseteq \mathcal{M}$ is such that for any $m_1, m_2 \in \mathcal{M}_\alpha$, there is no edge in the interference graph,  i.e.,  $(m_1, m_2) \notin \mathcal{E}_\alpha$. We follow the recursive largest first (RLF) algorithm \cite{leighton1979graph} to find the independent sets from the interference graph $\mathcal{G}_\alpha$. We choose $\mathcal{M}_\alpha^{(t)}$ as the $(t \mod I)$-th independent set, where $I$ is the number of independent sets returned by RLF. Overall procedure of $\alpha$-PMQ is summarized in Algorithm~\ref{alg:alpha_pmq}. 


\begin{algorithm}[t!] \label{alg:alpha_pmq}
    \caption{$\alpha$-PMQ}
    \SetKwInOut{Input}{Input}
    \Input{Codebook $\hat{\mathcal{X}}_{1:M}$; threshold $\alpha \in [0,1]$; channel matrix $H$, precoding matrix $W$, input symbol vector $s$}
    \SetKwInOut{Output}{Output}
    \Output{Discrete bits $b_{1:M}$}
    \textbf{Initialize} $\Delta_{m}^{(0)}=0_N$ for all $m=1,...,M$ ($0_N:$ $N\times 1$ vector of zeros); set $\mathcal{M}=\{1,...,M\}$\\
    Build an interference graph $\mathcal{G}_\alpha$ by following \eqref{eq:interference_graph_1}--\eqref{eq:interference_graph_3}

    Find disjoint collection of independent sets $\{\mathcal{M}_\alpha^i\}_{i=1}^I$ with $\cup_{i=1}^I \mathcal{M}_\alpha^{i}=\mathcal{M}$ using some heuristic algorithm (e.g., RLF)

    \For{\emph{$t=1,...,T$}}{
        determine independent nodes $\mathcal{M}_\alpha^{(t)} = \mathcal{M}_\alpha^{t \mod I}$ \\ 
        \For{\emph{$m \in \mathcal{M}_\alpha^{(t)}$}}{
        parallel search for independent nodes
        \begin{align} \nonumber
           \hspace{-0.38cm} b_{m}^{(t)} = \argmin_{ b_m \in \{0,1\}^{B_m}  } \big|\big|\sum_{m' \neq m} \Delta_{m'}^{(t-1)} + \Delta_m(  f(b_m|\hat{\mathcal{X}}_m)  )\big|\big|^2
        \end{align}
    update $\Delta_m^{(t)}$ using \eqref{eq:alpha_pmq_Delta_update}
    }
    \For{\emph{$m \notin \mathcal{M}_\alpha^{(t)}$}}{
    $\Delta_m^{(t)} = \Delta_m^{(t-1)}$ \\
    $b_m^{(t)} = b_m^{(t-1)}$
    }
    }
    \textbf{Return} $b^{(T)}_{1:M}$ \\
    \end{algorithm}

\subsection{Codebook Optimization with $\alpha$-PMQ} \label{subsec:codebookopt_alpha_pmq}
Given the algorithmic mapping $Q^\text{$\alpha$-PMQ}(\cdot)$ in (\ref{eq:local_quant})  between the codebook $\hat{\mathcal{X}}_{1:M}$ to the codewords $\hat{x}$ implemented by $\alpha$-PMQ, codebook optimization 
 is addressed via problem (\ref{eq:obj_wonju_codebook}).

\section{Neural Multivariate Quantization}\label{sec:neural_MQ}
All schemes presented so far require exhaustive searches over discrete spaces of cardinality growing exponentially with the fronthaul capacity. While this is feasible when the fronthaul capacity $B_m$ are, say, up to $12$--$13$ bits per channel use, it becomes quickly impractical for larger values of the fronthaul capacity $B_m$.  In this section, we introduce neural-network-based multivariate quantization, referred to as \emph{neural multivariate quantization} (neural-MQ), which replaces exhaustive search with gradient-based updates with the aid of a neural-network-based decoder. The approach is presented for general multi-antenna systems, and a modification of the EI criterion to this setting is also introduced in this section.

\begin{figure}[h]
  \begin{center}\includegraphics[scale=0.45]{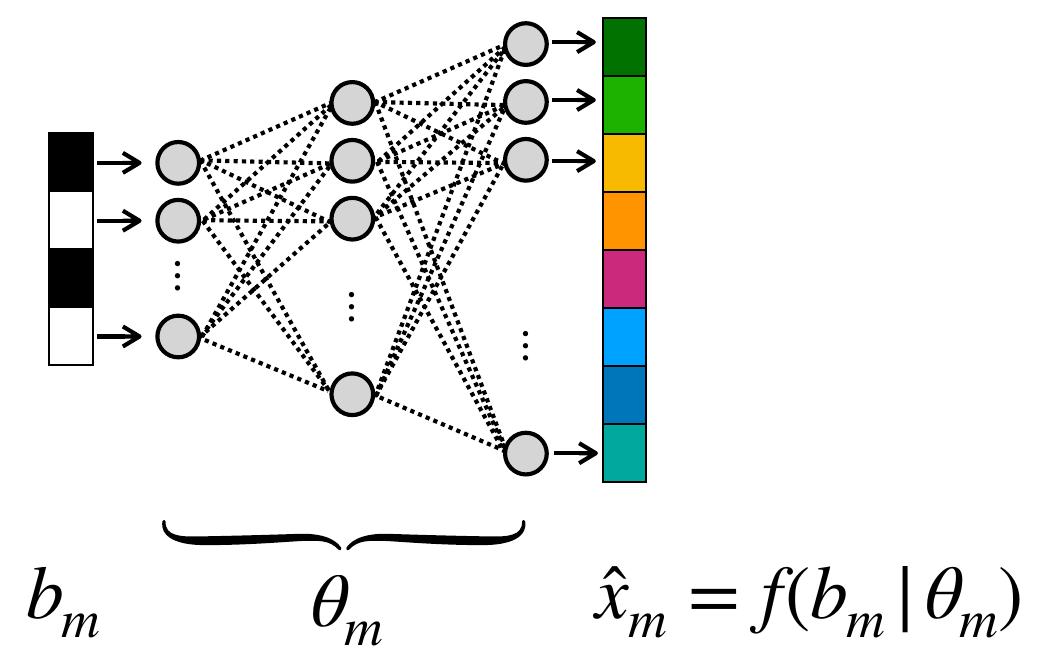}
    \end{center}
    \caption{ Illustration of neural codebook $f(b_m|\theta_m)$ {\color{black} for RU $m$}: the input is the binary message $b_m \in \{0,1\}^{B_m}$ of size $B_m$; the output is the transmitted symbol vector $\hat{x}_m \in \mathbbm{R}^{2  N_m^\text{tx}}$ of size $2N_m^\text{tx}$ with first $N_m^\text{tx}$ elements used for the real part of $\hat{x}_m$ while the remaining $N_m^\text{tx}$ elements for the imaginary part of $\hat{x}_m$. The number $K_m$ of hidden layers and the corresponding number $D_m$ of hidden neurons can be freely chosen, and we set by default $K_m=1$ and $D_m=B_m$.  }  
    \label{fig:neural_codebook_illustration}
 \end{figure}

\subsection{Neural Multivariate Quantization}
\label{subsec:NN_quan_dequan}
Instead of solving a discrete optimization problem of complexity $\mathcal{O}(2^{\sum_{m=1}^M B_m})$ like MQ, neural-MQ relaxes the search space to the vector space $\mathbb{R}^{\sum_{m=1}^M B_m}$, enabling the use of \emph{gradient-based updates}. This approach is used also in the literature on combinatorial optimization \cite{alet2018modular, nikoloska2022modular}. Following \cite{nikoloska2022modular}, we also incorporate an annealing mechanism to mitigate the distinction caused by the relaxation.


Specifically, as shown in Fig.~\ref{fig:neural_codebook_illustration}, neural-MQ introduces a neural network, or \emph{neural codebook}, $f(b_m|\theta_m)$, parameterized by a vector $\theta$ that determines the synaptic weights and biases of the neural network \cite{simeone2022machine},  that takes as an input the binary message $b_m \in \{0,1\}^{B_m}$  for RU $m$ and outputs the corresponding signal $\hat{x}_m \in \mathbb{C}^{N_m^\text{tx}}$ for each RU $m=1,...,M$. Note that in the previous sections, the function $f_m(\cdot)$ was implemented by a look-up table $\hat{\mathcal{X}}_{m}$ that recovers the quantized signal $\hat{x}_m$ given the index $b_m$.

Accordingly, quantization is carried out, as in (\ref{eq:obj_MQ_sota}), by addressing the problem 
\begin{align}
    \label{eq:N_MQ_obj}  b_{1:M}^\text{neural-MQ} = \argmin_{ b_{1:M} \in \{0,1\}^{\sum_{m=1}^M B_m}} \sum_{n=1}^N \text{EI}_n ( x, f(b_{1:M}|\theta_{1:M}) ),
\end{align}
{\color{black} from which we have the overall quantized vector}
\begin{align}
    \hat{x} &= Q^\text{neural-MQ}(x|\theta_{1:M}) = f(b_{1:M}^\text{neural-MQ}|\theta_{1:M}),
\end{align}
{\color{black} given } $f(b_{1:M}|\theta_{1:M})=[ f(b_1|\theta_1)^\top,..., f(b_M|\theta_M)^\top ]^\top$.

To address the combinatorial problem (\ref{eq:N_MQ_obj}), we relax the problem  by considering the optimization 
\begin{align} \label{eq:gradient_based_quan}
    r_{1:M} = \argmin_{r_{1:M}\in \mathbb{R}^{\sum_{m=1}^M B_m}} \sum_{n=1}^N \text{EI}_n( x, f(\sigma_\tau(r_{1:M}) |\theta_{1:M} ))
\end{align}
over the real-valued vector $r_{1:M}$, where we have defined the sigmoid function with temperature $\tau$ as \cite{simeone2022machine}
\begin{align} \label{eq:sigmoid}
    \sigma_\tau(x) = \frac{1}{1+\exp{(-x/\tau)}},
\end{align}
which is applied element-wise in (\ref{eq:gradient_based_quan}). The sigmoid function (\ref{eq:sigmoid})  squashes (unconstrained) real number into an interval $[0,1]$. As the temperature $\tau$ decreases, the sigmoid function $\sigma_\tau(\cdot)$ becomes increasingly close to a hard step function.

Accordingly, given an optimized real vector $r_{1:M}$, the final solution is obtained as 
\begin{align} \label{eq:neural_hard_b}
    b_{1:M}= \sigma_{0}(r_{1:M}) = \mathbbm{1}(r_{1:M}>0),
\end{align}
which can then be transmitted to each RU $m$ via the fronthaul link of capacity $B_m$. After receiving $b^m$, each RU $m$ can transmit the signal by simply running the neural network $f(\cdot|\theta_m)$ once, i.e., computing $\hat{x}_m = f(b_m|\theta_m)$ and transmitting $\tilde{x}_m=\gamma_m \hat{x}_m$ with power scaling factor $\gamma_m$.

Problem (\ref{eq:gradient_based_quan}) is addressed via gradient descent (GD) with temperature annealing \cite{nikoloska2022modular}. Thus,  we decrease temperature by setting the temperature at $i$-th GD step as $\tau_i = \exp(-5 \cdot i/ I) $ for a given total number $I$ of GD updates. The overall procedure of neural-MQ is summarized in Algorithm~\ref{alg:neural_MQ}. 

The computational complexity of neural-MQ is determined by the number $I$ of GD updates as well as the number of (hidden) neurons in the neural network $f(\cdot|\theta_m)$.  For instance if one adopts $f(\cdot|\theta_m)$ as a fully-connected neural network with $K_m \geq 1$ hidden layers each with  $D_m$ neurons, solving (\ref{eq:N_MQ_obj}) via (\ref{eq:gradient_based_quan}) would require computational complexity of the order $\mathcal{O}(2 I\cdot \sum_{m=1}^M  D_m(B_m + D_m(K_m-1) + 2N_m^\text{tx} )$ \cite{simeone2022machine, griewank1993some} which grows \emph{linearly} with fronthaul sum-rate $\sum_{m=1}^M B_m$  at the DU side, while at the RU side, it requires $\mathcal{O}( D_m(B_m + D_m(K_m-1) + 2N_m^\text{tx}))$ for each RU $m$ (see Table~\ref{table:tab_comp_comp}).



\subsection{Multi-Antenna Effective Interference}
As explained in the previous section, the EI criterion \eqref{eq:EI_def} studied in \cite{lee2016multivariate} and reviewed in Sec.~\ref{sec:wonju} assumes single-antenna for both RUs and UEs. In this subsection, we extend, and modify, the notion of EI to the scenario presented in Sec.~\ref{sec:sys_model} in which both RUs and UEs are equipped with multiple antennas.  As in the previous section, we simplify the notation by removing the explicit dependence on the channel use index $k$.

Accounting for the effect of the beamforming receiver matrix, the expected discrepancy between the desired symbol $s_n$ and the estimated symbol $\hat{s}_n$ can be characterized by the $N_n^\text{rx} \times N_n^\text{rx}$ \emph{error covariance matrix} (see, e.g., \cite{christensen2008weighted})
\begin{align} \label{eq:error_cov}
    E_n &= \mathbb{E}[(\hat{s}_n - s_n)(\hat{s}_n -s_n )^\dagger] \nonumber\\&= F_n^\dagger H_n \mathbb{E}[\tilde{x}\tilde{x}^\dagger]H_n^\dagger F_n - F_n^\dagger H_n \mathbb{E}[\tilde{x}s_n^\dagger]  + \sigma^2 F_n^\dagger F_n - \mathbb{E}[s_n \tilde{x}^\dagger] H_n^\dagger F_n + I,
\end{align}
where the expectation is taken with respect to the random symbol as well as  the noise vector in \eqref{eq:rx_basic}. We propose to use as EI criterion the trace of the error covariance matrix, which can be simplified as follows by retaining only the terms that depend on the quantization process
\begin{align} \label{eq:new_EI}
    \text{EI}_n(x,\hat{x}) = \text{Re}\big( \text{tr}(F_n^\dagger H_n \Gamma \hat{x}(F_n^\dagger H_n \Gamma \hat{x}- 2 s_n)^\dagger)\big).
\end{align} Note that the precoded signal $x$ appears implicitly in (\ref{eq:new_EI}), as for the original EI (\ref{eq:EI_def}), through its individual precoded signals. In particular, the EI criterion (\ref{eq:new_EI}) recovers (\ref{eq:EI_def})  by setting $F_n=((H_nW_n)^\dagger)^{-1}$.

\subsection{Receive Beamforming} \label{subsec:EI_with_CE}
\label{subsec:ch_est}
The EI criterion (\ref{eq:new_EI}) is defined for any receive beamforming matrix $F_n$. For instance, MMSE receiver can be adopted 
\cite{feng2021weighted}
\begin{align} \label{eq:MMSE_receiver_beamforming}
    F_n = \bigg( \sum_{n'=1}^N H_n W_{n'} W_{n'}^\dagger H_{n}^\dagger  + \sigma^2 I \bigg)^{-1} H_n W_n,
\end{align}
which requires the availability of  all the  effective channels {\color{black} $\{ H_n W_{n'} \}_{n'=1}^N$, that is, of its own effective channel $H_n W_n$ as well as of the effective channels for all the other  UEs $n'\neq n$}. A more practical approach would be to adopt a localized MMSE receiver, which can be justified by the channel hardening effect of cell-free massive MIMO (see, e.g., \cite{ngo2024ultra, gholami2022survey}), i.e., 
\begin{align} \label{eq:simpler_MMSE_receiver_beamforming}
    F_n = \Big(  H_n W_{n} W_{n}^\dagger H_{n}^\dagger  + \sigma^2 I \Big)^{-1} H_n W_n.
\end{align}

\subsection{Neural Codebook Optimization} \label{subsec:codebook_opt_neural_MQ}
The parameter vector $\theta_{1:M}$ of neural codebook $f(\cdot|\theta_m)$ for $m=1,...,M$ in Sec.~\ref{subsec:NN_quan_dequan} can be optimized in a similar manner to (\ref{eq:obj_wonju_codebook}), i.e., 
\begin{align}
    \label{eq:obj_neural_codebook}
    \min_{\theta_{1:M}} \mathbbm{E}\Bigg[ \sum_{n=1}^N \text{EI}_n(x, \hat{x}) \Bigg],
\end{align}
where $\hat{x}$ is obtained from (\ref{eq:N_MQ_obj}),  and  the average is taken over both symbols $s$ and channels $H$.

\subsection{Extending $\alpha$-PMQ to Multi-Antenna Systems} \label{subsec:alpha_pmq_for_multi_antenna}
As a final note, using (\ref{eq:new_EI}), it is straightforward to extend $\alpha$-PMQ, presented in Sec.~\ref{sec:PMQ}, to multi-antenna systems by noting that EI in (\ref{eq:new_EI}) can be also decomposed into the terms that explicitly shows the dependence on each RU $m$. This can be done as
\begin{align} \label{eq:new_EI_decomp}
    &\text{EI}_n(x,\hat{x}) = \text{Re}\bigg( \text{tr}\bigg( \Big[\sum_{m=1}^M {F_{n,m}}^\dagger H_{n,m} \hat{x}_m \Big] \Big[ \Big( \sum_{m'=1}^M F_{n,m'}^\dagger H_{n,m'} \hat{x}_{m'}\Big)^\dagger  2 s_n^\dagger   \Big] \bigg)\bigg),
\end{align}
denoting as $F_{n,m}$ and $H_{n,m}$ the beamforming matrix of UE $n$ designed for  RU  $m$ and the channel matrix between UE $n$ and RU $m$, respectively.

\begin{algorithm}[t!] \label{alg:neural_MQ}
    \caption{Neural-MQ}
    \SetKwInOut{Input}{Input}
    \Input{Number of iterations $I$; neural codebook $f_m(\cdot)$ for $m=1,...,M$, symbol vector $s$; channel matrix $H=[H_1^\top, ..., H_N^\top]^\top$; precoding matrix $W=[W_1,...,W_N]$; receive beamforming matrix $F=[F_1, ..., F_N]$; step size $\eta > 0$ }
    \SetKwInOut{Output}{Output}
    \Output{Discrete bits $b_{1:M}$}
    \textbf{Initialize} $r_{1:M}$ as the all-zero vectors; $\tau_1 = 1$
    
    \For{\emph{$i=1,...,I$}}{
    \begin{align} \label{eq:GD_for_neural_MQ}
        r_{1:M} &\leftarrow   r_{1:M} \\&- \eta \cdot \nabla_{r_{1:M}} \sum_{n=1}^N \text{EI}_n(x, f(\sigma_{\tau_i}(r_{1:M})|\theta_{1:M}))    \nonumber
    \end{align}
    update sigmoid temperature $\tau_i = \exp(-5 \cdot i/ I) $
    }

    obtain discrete binary messages $b_{1:M}$ in (\ref{eq:neural_hard_b})
    
    \textbf{Return} $ b_{1:M}$ \\
    \end{algorithm}

\section{Experimental Results{\protect\footnote{Code is available at https://github.com/kclip/scalable-MQ.}}} \label{sec:experiments}
In this section, we investigate the scalability of the proposed PC-based solutions by separating the analysis into  low-fronthaul and high-fronthaul regimes. In the low fronthaul regime, we compare the proposed $\alpha$-PMQ (Sec.~\ref{sec:PMQ})  and neural-MQ (Sec.~\ref{sec:neural_MQ})  against conventional PC schemes, i.e., {\color{black} VQ (Sec.~\ref{sec:VQ}) \cite{lee2016multivariate}}, SMQ (Sec.~\ref{sec:smq}), and MQ (Sec.~\ref{subsec:MQ_given_codebook}). {\color{black} We specifically focus on a regime with a  low fronthaul capacity, in which CP cannot be evaluated even under the minimum scalar quantization level $B^\text{CP}=1$ with maximum precoding sharing factor $K^\text{CP} = 4\cdot 12\cdot 14$ \cite{grönland2023learningbased}. }

In the high-fronthaul regime, we focus on comparing neural-MQ with CP, since all the other PC-based solutions, including $\alpha$-PMQ, cannot be run in a reasonable time due to their exponential complexity.  {\color{black} In this regime}, we consider instead two \emph{infinite-fronthaul benchmarks}. The first stipulates that the RUs transmit directly the linearly precoded vector $x=Ws$. 

In the second, we address problem (\ref{eq:gradient_based_quan}) by optimizing the $N^\text{tx}$ transmitted signal $\hat{x}$ as $\hat{x} = \argmin_{\hat{x} \in \mathbb{C}^{N^\text{tx}}} \\ \sum_{n=1}^N\text{EI}_n(x,\hat{x})$, where $\hat{x}$ is an arbitrary complex-valued vector. This solution  can be considered to be a type of \emph{non-linear precoding}, which may outperform linear precoding \cite{schmidt2008minimum}.   Being based on optimization (\ref{eq:gradient_based_quan}), the performance of the proposed neural-MQ scheme is upper bounded by the second, non-linear, infinite-fronthaul benchmark scheme, but not by the first, linear precoding, benchmark.


For all the experiments, we set the  total signal-to-noise ratio (SNR) as $ \sum_{m=1}^M P_m/\sigma^2 = 18$ dB, unless specified otherwise; and we assume the same power $P_m$ for all the RUs.  Codebook optimization for both $\alpha$-PMQ (Sec.~\ref{subsec:codebookopt_alpha_pmq}) and neural-MQ (Sec.~\ref{subsec:codebook_opt_neural_MQ})  is done by assuming Gaussian information symbols using $100$ different channel realizations and $100$ different Gaussian symbols. {\color{black} Specifically, we use \texttt{CVX} \cite{gb08} for codebook optimization of $\alpha$-PMQ, while the Adam optimizer  \cite{kingma2014adam} is adopted for codebook optimization of neural-MQ.  }

{\color{black} The performance of the PC-based solutions are evaluated based on the sum-spectral efficiency and the symbol error rate.}
The sum-spectral efficiency is evaluated as $\mathbb{E}_H[ \sum_{n=1}^N \log \det ( I + \mathbb{E}_s[E_n] ) - \log \det (\mathbb{E}_s[E_n] )]$, where the expectation over symbols $\mathbb{E}_s[\cdot]$  is estimated by $1000$ Gaussian symbols, while the expectation over channels $\mathbb{E}_H[\cdot]$ is estimated by five different channel realizations following 3GPP urban micro (UMi) channel models \cite{3gpp}.  When evaluating symbol error rate, we consider 16 quadrature amplitude modulation (16-QAM) symbols, and report the average symbol error rate over the $N$ UEs.

{\color{black} Lastly, w}e adopt MMSE precoding  and localized MMSE receive beamforming (\ref{eq:simpler_MMSE_receiver_beamforming}). For neural-MQ, we set the number of GD steps $I=100$, number of hidden layers $K=1$, and the number of hidden neurons $D_m=B_m$,  unless specified otherwise.

 \begin{figure}[t]
  \begin{center}\includegraphics[scale=0.17]{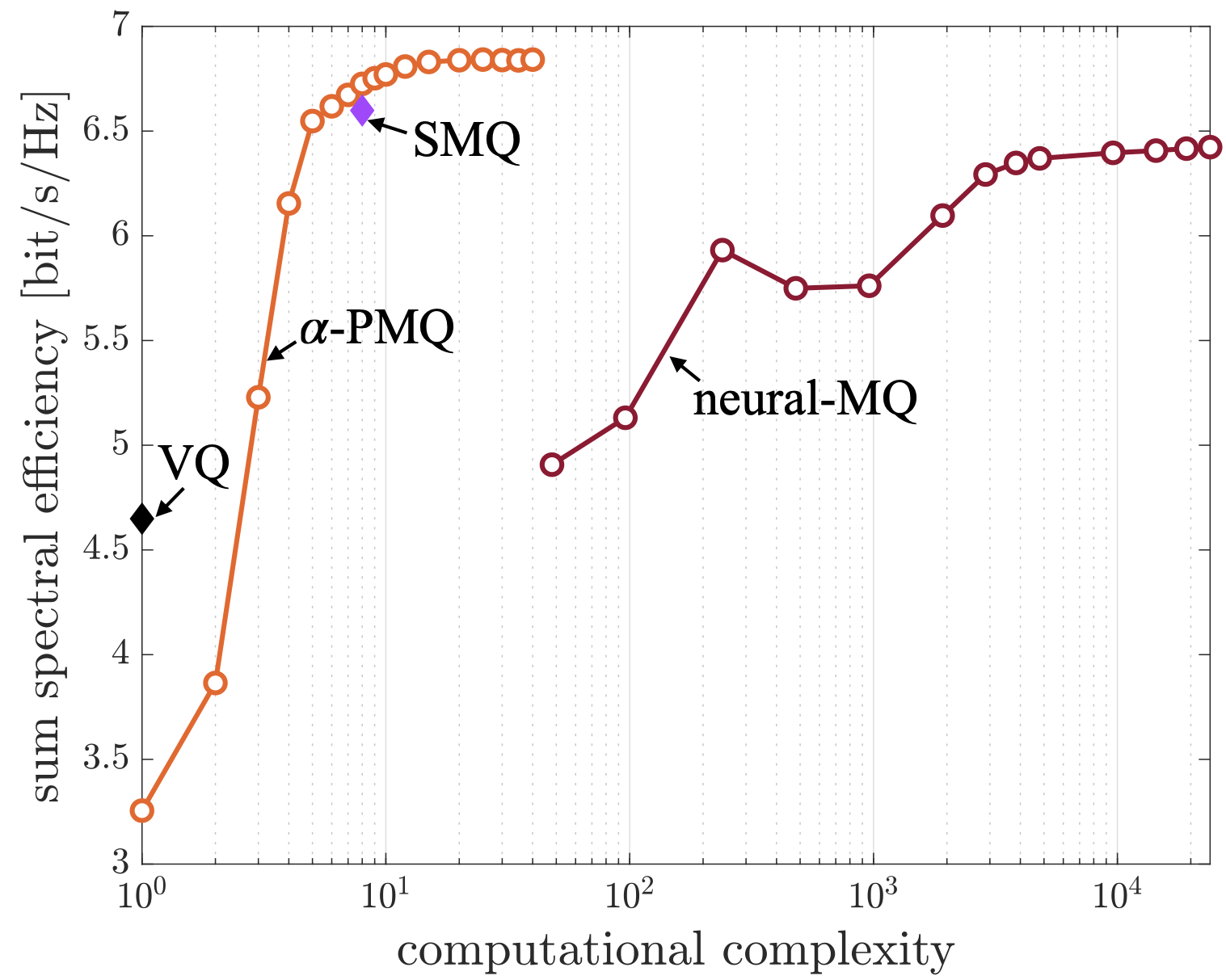}
    \end{center}
    \caption{Spectral efficiency as a function of the computational complexity {\color{black}normalized with respect to the computational complexity required by VQ}. We assume  $M=8$ RUs equipped with $N_m^\text{tx}=2$ antennas under fronthaul capacity $B_m=2$ and $N=3$  UEs equipped with $N_n^\text{rx}=2$ antennas that wish to receive $L_n=1$ data stream.  The positions of the RUs are  fixed, while UE positions are randomly distributed for each channel realization. The channel is generated by following 3GPP urban micro (UMi) models \cite{3gpp}.}
    \label{fig:low_FH}
 \end{figure}

\subsection{Low-Fronthaul Capacity: Computational Complexity Analysis}
First, we study the spectral efficiency attained by different PC-based fronthaul quantization schemes against the computational complexity of fronthaul quantization. To this end, we consider the number of operations on a parallel processor, whereby parallel operation at different subprocessors are counted only once. {\color{black} We recall that Table~\ref{table:tab_comp_comp} reviews the complexity measures for all schemes}. 


In Fig.~\ref{fig:low_FH}, we assume $N=3$ multi-antenna UEs equipped with $N_n^\text{rx}=2$ antennas with $L_n=1$  data stream per each UE, served by $M=8$ RUs equipped with $N_m^\text{tx}=2$ antennas with fronthaul capacity $B_m=2$.  We set $\alpha=0.5$ for $\alpha$-PMQ, and we vary the complexity of $\alpha$-PMQ and neural-MQ by increasing the number $T$ of iterations (Algorithm~\ref{alg:alpha_pmq}) and the number $I$ of GD steps (Algorithm~\ref{alg:neural_MQ}), respectively.  Accordingly, all the other schemes provide a single point in the performance-complexity trade-off.   MQ is not shown in the figures due to excessive computational complexity.

It can be seen from the figure that, in this low-fronthaul capacity regime, for all spectral efficiency larger than around $4.75$ bit/s/Hz,  $\alpha$-PMQ  outperforms the other schemes, only VQ can achieve a higher computational efficiency but only in the very low spectral efficiency regime.

\begin{figure*}[t]
  \begin{center}\includegraphics[scale=0.17]{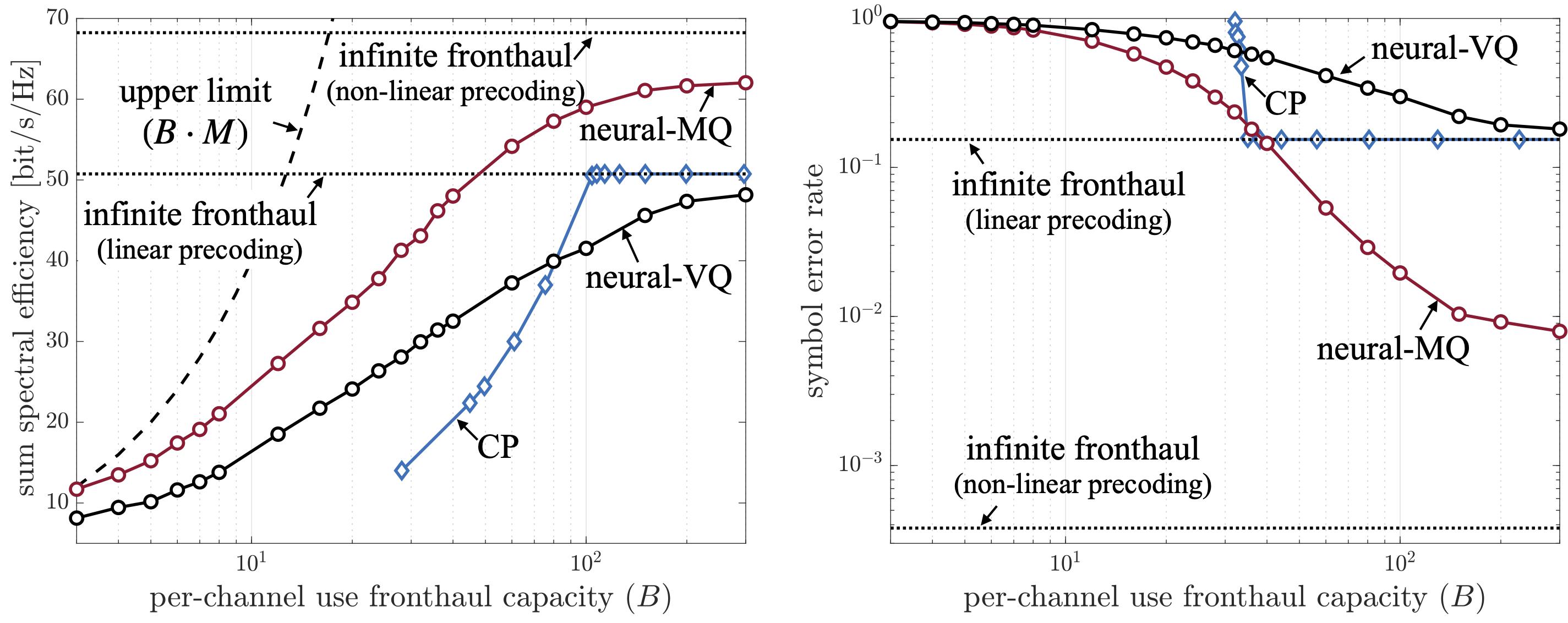}
    \end{center}
    \caption{Sum spectral efficiency (left) and averaged symbol error rate (right) as a function of per-channel use fronthaul capacity $B$. We assume  $M=4$ RUs equipped with $N_m^\text{tx}=16$ antennas and $N=4$ multi-antenna UEs equipped with $N_n^\text{rx}=4$ antennas that wishes to receive $L_n=2$ independent data streams. The positions of the RUs are fixed, while UE positions are randomly distributed for each channel realization. The channel is generated by following 3GPP urban micro (UMi) models \cite{3gpp}. We assume $16$-QAM for symbol error rate evaluation. }
    \label{fig:per_B_NN}
 \end{figure*}

\subsection{Higher-Fronthaul Capacity Regime: Comparison Between CP and PC} \label{subsec:per_B}

In this subsection, {\color{black} focusing on the high-fronthaul capacity regime,}  we compare neural-MQ to (\emph{i}) infinite-fronthaul benchmarks; (\emph{ii}) CP; and  (\emph{iii}) neural-VQ (see Appendix~\ref{app:neural_VQ}).

For CP, we consider four different standard levels of precoding sharing, namely $K^\text{CP}=48\cdot 12\cdot 14, 36\cdot 12\cdot 14, 24\cdot 12\cdot 14, 12\cdot 12\cdot 14$,  while fixing the quantization level to $B^\text{CP}=16$ \cite{grönland2023learningbased}. In order to further examine the behavior of CP,  we also considered the non-standard choices $B^\text{CP}=2,4$ bits for  $K^\text{CP}=48\cdot 12\cdot 14$, and $B^\text{CP} =32,64,128,256,512,1024,2048$ for  $K^\text{CP}=12\cdot 12\cdot 14$. We design the precoding matrix for CP  by first {\color{black} evaluating the averaged channel matrix across $K^\text{CP}$ channel uses,  and  then applying the MMSE precoding scheme \cite{vu2007mimo, venugopal2019optimal}}.

In Fig.~\ref{fig:per_B_NN}, we show both sum spectral efficiency (left) and symbol error rate (right).  For the fronthaul overhead (\ref{eq:CP_constraint}) of CP, we assume channel coding with rate $R_\text{code}=0.5$. This choice is dictated by the fact that a modulation scheme with $2^M$ symbols is approximately sufficient to achieve capacity $M/2$ \cite{le2003signal}.

Neural-MQ is seen to outperform CP in all fronthaul capacity regimes, while CP outperforms neural-VQ in the regime of $B_m \geq 80$, which highlights the importance of MQ. Note that both neural-VQ and CP cannot outperform the linear precoding benchmark, unlike the proposed neural-MQ.

Furthermore,  neural-MQ approaches the performance of the  infinite-fronthaul capacity with non-linear precoding benchmark as the fronthaul capacity increases, while neural-VQ approaches the infinite-fronthaul capacity benchmark under linear precoding. This  validates the effectiveness of the proposed sum-EI criterion (\ref{eq:new_EI}) in ensuring interference management, as well as of the proposed neural-codebook design, which can leverage the benefits of non-linear precoding.


\subsection{Impact of Number of Gradient Descent Steps for Neural-MQ}

To provide further insights into the design of neural-MQ, we now study the impact of number $I$ of gradient descent steps in (\ref{eq:GD_for_neural_MQ}) on the performance of neural-MQ. We further compare the performance over different architectures of the neural codebook by varying the number of hidden layers $K=0,1,2$. Note that $K=0$ corresponds to a linear mapping  (look-up table) from bits to codewords. 

\begin{figure}[t]
  \begin{center}\includegraphics[scale=0.17]{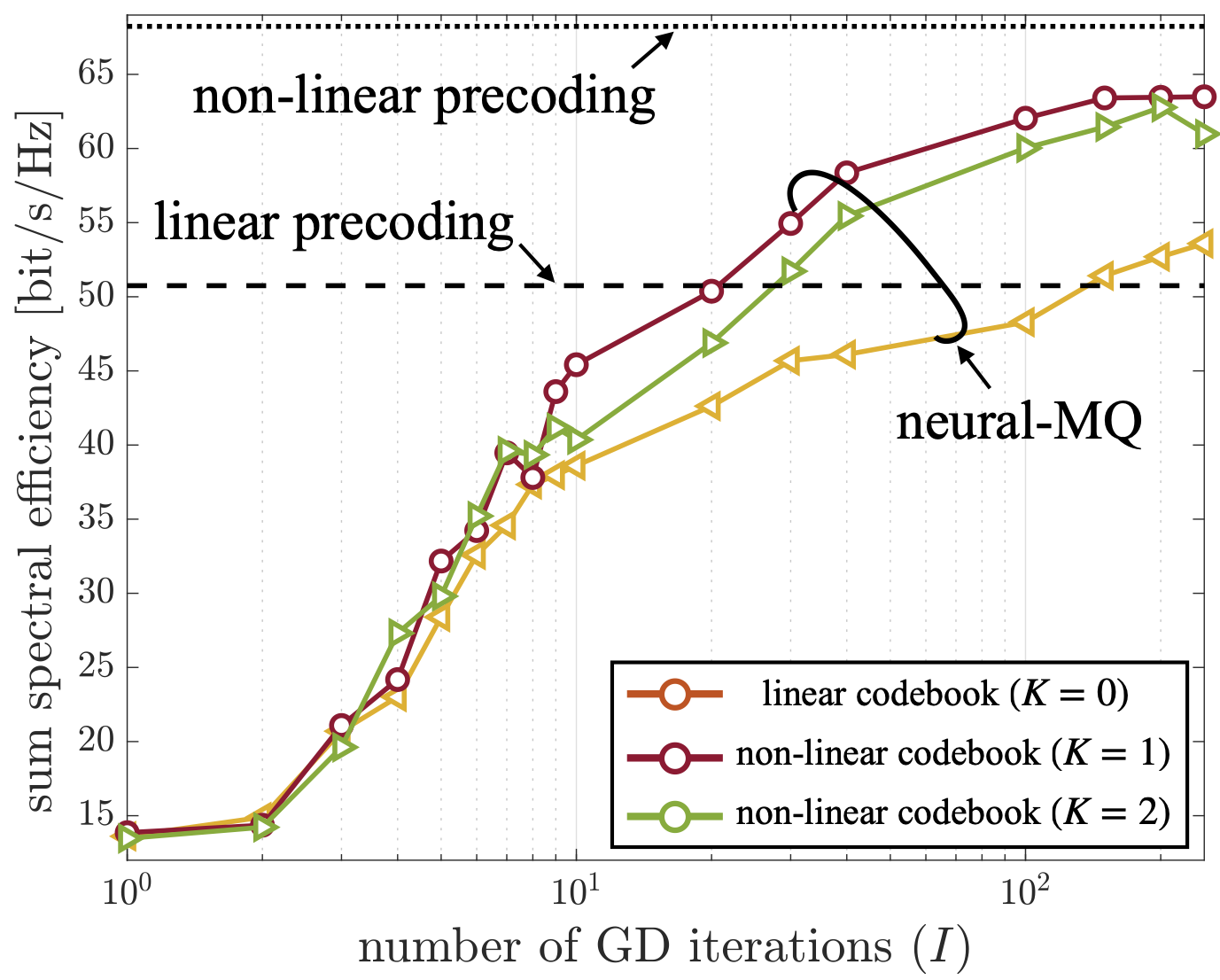}
    \end{center}
    \caption{  Sum spectral efficiency as a function of total number $I$ of gradient descent (GD) iterations used by neural-MQ at the DU side.   We set $B_m=300$.  Other settings are the same as in Fig.~\ref{fig:per_B_NN}. }
    \label{fig:per_iter}
 \end{figure}

Fig.~\ref{fig:per_iter} shows that, under a sufficiently large fronthaul capacity, here $B_m=300$, and  number of GD iterations ($I \geq 200$), neural-MQ outperforms linear precoding with infinite-fronthaul capacity, even with a linear mapping (look-up table) $(K=0)$. Furthermore, a single hidden layer $(K=1)$ is observed to perform the best, outperforming the linear precoding with infinite-fronthaul capacity with less number of GD iterations ($I\geq 20$).

\subsection{Generalization Capacity of Neural-MQ}
\begin{figure}[t]
  \begin{center}\includegraphics[scale=0.17]{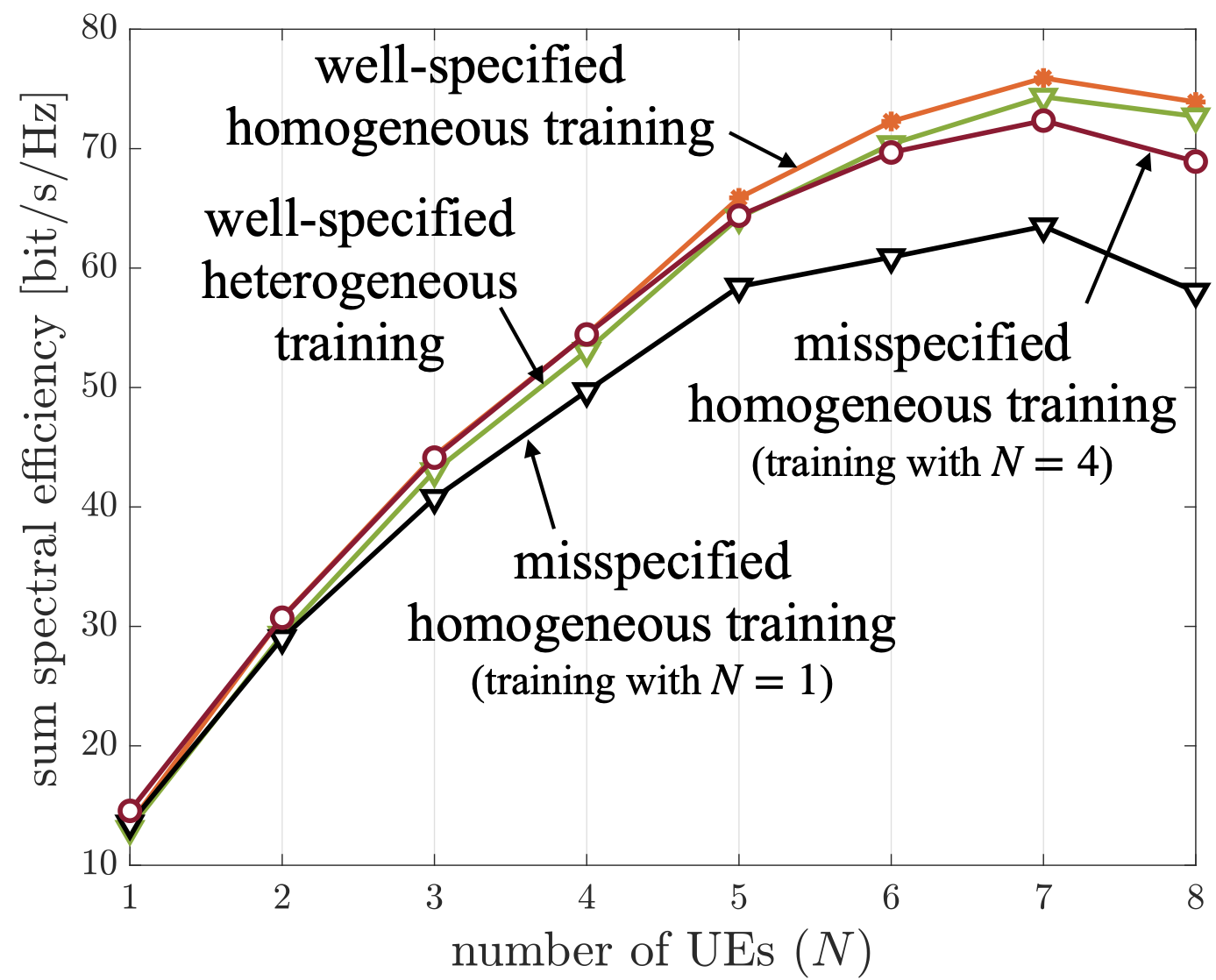}
    \end{center}
    \caption{Sum spectral efficiency as a function of number $N$ of UEs during testing. We set $B_m=64$. Other settings are the same as in Fig.~\ref{fig:per_B_NN}.  }
    \label{fig:per_N_NN}
 \end{figure}

We now investigate the generalization capacity of neural-MQ by changing the number $N$ of UEs between training and testing. Specifically, in Fig.~\ref{fig:per_N_NN}, we consider the following scenarios: (\emph{i}) well-specified homogeneous training condition, in which we have no discrepancy between the fixed number $N$ of UEs during training and testing; (\emph{ii}) misspecified homogeneous training conditions,  in which training assumes a  number $N$ of UEs equal to either $1$ or $4$, while testing for $N=1,...,8$ UEs; (\emph{iii}) well-specified heterogeneous training conditions, in which we train and test neural-MQ by using data set that consists of the same number $N=1,...,8$ of UEs. 

Well-specified homogeneous training is seen to yield the best performance,  with similar performance achieved also with well-specified heterogeneous training. While misspecified homogeneous training that assumes a single UE performs poorly when the number of UEs increases ($N \geq 5$), assuming $N=4$ UEs during training is sufficient to obtain good performance when testing with either smaller or larger numbers of UEs. The observed generalization capacity of neural-MQ can be understood by noting that the main role of neural-MQ is interference management, which depends less strongly on exact number $N$ of UEs and more on aspects such as the placement of RUs.

\begin{figure}[h]
  \begin{center}\includegraphics[scale=0.17]{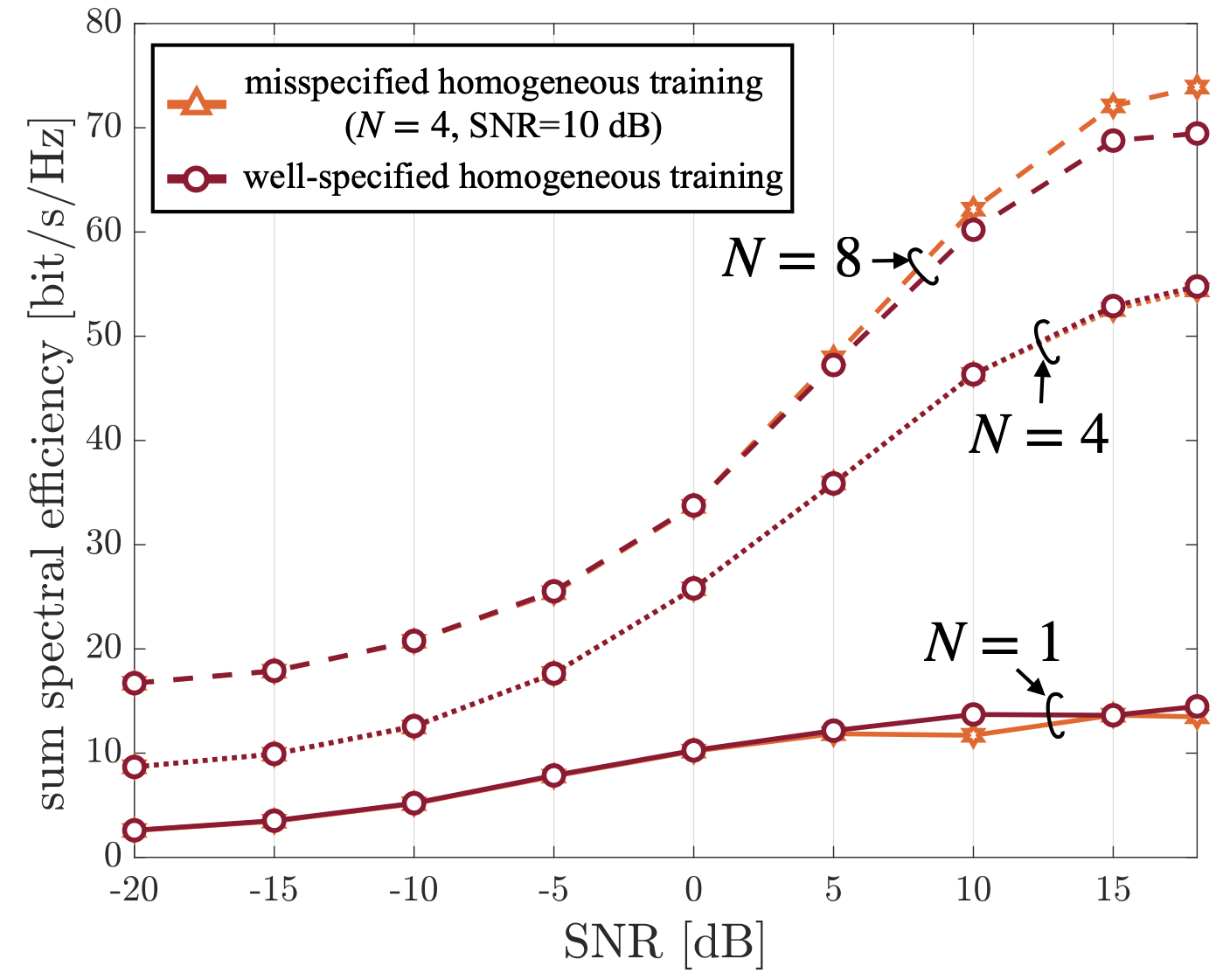}
    \end{center}
    \caption{ Sum spectral efficiency  as a function of SNR $\sum_{m=1}^M P_m/\sigma^2$ during testing, for different numbers $N=1,4,8$ of UEs.  We set $B_m=64$. Other settings are the same as in Fig.~\ref{fig:per_B_NN}.  }
    \label{fig:per_N_NN_SNR}
 \end{figure}

Next, we study the generalization capacity of neural-MQ by changing the total SNR $ \sum_{m=1}^M P_m/\sigma^2$ from $-10$ dB to $18$ dB. In a manner similar to Fig.~\ref{fig:per_N_NN}, we consider (\emph{i}) well-specified homogeneous training, i.e., no discrepancy between training and testing for both number $N$ of UEs as well as SNR, and (\emph{ii}) misspecified homogeneous training in which training is done only by assuming $N=4$ and $\sum_{m=1}^M P_m/\sigma^2 = 10$ dB. It can be concluded from Fig.~\ref{fig:per_N_NN_SNR} that the considered misspecified homogeneous training is sufficient to obtain performance comparable to the well-specified homogeneous case. Robustness to different SNRs is again attributed to the main task of neural-MQ being interference management. 

\section{Conclusion} \label{sec:conclusion}
In this paper, we have proposed  scalable MQ solutions for PC strategies in cell-free massive MIMO.  The first approach, $\alpha$-PMQ, which has computational complexity  growing exponentially in the per-RU  fronthaul rate, outperforms all the other benchmarks including VQ  \cite{lee2016multivariate} (e.g., $\times 1.5$ sum spectral efficiency, with marginally increased complexity) and SMQ  \cite{lee2016multivariate} (e.g., $\times 1.05$ sum spectral efficiency, with slightly increased complexity),  under conditions in which the original MQ \cite{lee2016multivariate} becomes infeasible due to large number of RUs (e.g., $M=8$). 

The second proposed approach, neural-MQ, has computational complexity that grows linearly in the fronthaul sum-rate, and hence it can be applied in higher per-RU fronthaul regimes for which $\alpha$-PMQ is not feasible. Neural-MQ was seen to always outperform CP, irrespective of the fronthaul capacity (up to $\times 2.5$ sum spectral efficiency under the same fronthaul capacity).

Interesting future work direction may include  reducing the computational complexity of neural-MQ at the DU side by enabling more efficient GD steps possibly aided by adaptive learning rate \cite{duchi2011adaptive}, by deep unfolding \cite{shlezinger2022model}, and/or by a  meta-learned transformers \cite{jain2024mnemosyne}.

\appendices
\section{Neural Vector Quantization} \label{app:neural_VQ}
Neural-VQ follows the same procedure as in neural-MQ while the only difference coming from the objective of the minimization  in (\ref{eq:N_MQ_obj}), i.e., 
\begin{align}
    \label{eq:N_VQ_obj}  b_{m}^\text{neural-VQ} = \argmin_{ b_{m} \in \{0,1\}^{B_m}} || x_m - f(b_m|\theta_m) ||^2,
\end{align}
from which we have the quantized vector for RU $m$ as 
\begin{align}
    \hat{x}_m &= Q^\text{neural-VQ}(x_m|\theta_{m}) = f(b_{m}^\text{neural-MQ}|\theta_{m}).
\end{align}
In a manner similar to neural-MQ, we relax the problem by considering the optimization 
\begin{align} \label{eq:gradient_based_quan_VQ}
    r_{m} = \arg\min_{r_{1:M}\in \mathbb{R}^{ B_m}}  || x_m -  f(\sigma_\tau(r_{m}) |\theta_{m} )||^2, 
\end{align}
which can again be solved via GD with temperature annealing as shown in Algorithm~\ref{alg:neural_VQ}. It can be easily checked that the computational complexity of neural-VQ, $\mathcal{O}(D^2)$, is less than VQ, $\mathcal{O}(2^{B_m})$, under high fronthaul capacity regime. 

\begin{algorithm}[t!] \label{alg:neural_VQ}
    \caption{Neural-VQ}
    \SetKwInOut{Input}{Input}
    \Input{Number of iterations $I$; neural codebook $f_m(\cdot)$ for DU $m$, symbol vector $s$; precoding matrix $W_m$ associated to RU $m$; step size $\eta > 0$ }
    \SetKwInOut{Output}{Output}
    \Output{Discrete bits $b_{m}$}
    \textbf{Initialize} $r_{m}$ as the all-zero vectors; $\tau_1 = 1$
    
    \For{\emph{$i=1,...,I$}}{
    \begin{align} \label{eq:GD_for_neural_VQ}
        r_{m} &\leftarrow   r_{m} - \eta \cdot \nabla_{r_{m}} || x_m - f(\sigma_{\tau_i}(r_{m})|\theta_{m})||^2 
    \end{align}
    update sigmoid temperature $\tau_i = \exp(-5 \cdot i/ I) $
    }

    obtain discrete binary message $b_{m} = \mathbbm{1}(r_m > 0)$
    
    \textbf{Return} $ b_{m}$ \\
    \end{algorithm}

\bibliographystyle{IEEEtran}
\bibliography{ref.bib}

\begin{thebibliography}{10}
\providecommand{\url}[1]{#1}
\csname url@samestyle\endcsname
\providecommand{\newblock}{\relax}
\providecommand{\bibinfo}[2]{#2}
\providecommand{\BIBentrySTDinterwordspacing}{\spaceskip=0pt\relax}
\providecommand{\BIBentryALTinterwordstretchfactor}{4}
\providecommand{\BIBentryALTinterwordspacing}{\spaceskip=\fontdimen2\font plus
\BIBentryALTinterwordstretchfactor\fontdimen3\font minus \fontdimen4\font\relax}
\providecommand{\BIBforeignlanguage}[2]{{%
\expandafter\ifx\csname l@#1\endcsname\relax
\typeout{** WARNING: IEEEtran.bst: No hyphenation pattern has been}%
\typeout{** loaded for the language `#1'. Using the pattern for}%
\typeout{** the default language instead.}%
\else
\language=\csname l@#1\endcsname
\fi
#2}}
\providecommand{\BIBdecl}{\relax}
\BIBdecl

\bibitem{simeone2012cooperative}
O.~Simeone, N.~Levy, A.~Sanderovich, O.~Somekh, B.~M. Zaidel, H.~V. Poor, and S.~Shamai, ``Cooperative wireless cellular systems: An information-theoretic view,'' \emph{Foundations and Trends{\textregistered} in Communications and Information Theory}, vol.~8, no. 1-2, pp. 1--177, 2012.

\bibitem{ngo2024ultra}
H.~Q. Ngo, G.~Interdonato, E.~G. Larsson, G.~Caire, and J.~G. Andrews, ``Ultra-dense cell-free massive {{MIMO}} for {6G}: Technical overview and open questions,'' \emph{arXiv preprint arXiv:2401.03898}, 2024.

\bibitem{rodriguez2020cloud}
V.~Q. Rodriguez, F.~Guillemin, A.~Ferrieux, and L.~Thomas, ``Cloud-ran functional split for an efficient fronthaul network,'' in \emph{Proc. IEEE International Wireless Communications and Mobile Computing (IWCMC)}, Limassol, Cyprus, June 2020.

\bibitem{grönland2023learningbased}
A.~Gr{\"o}nland, B.~Klaiqi, and X.~Gelabert, ``Learning-based latency-constrained fronthaul compression optimization in {C-RAN},'' in \emph{Proc. IEEE 28th International Workshop on Computer Aided Modeling and Design of Communication Links and Networks (CAMAD)}, Edinburgh, Scotland, Nov. 2023.

\bibitem{5g_oran_spectrum}
M.~Koziol, ``5{G} ``open'' standards are in play—just how open are they?'' in \emph{IEEE Spectrum}, 2023.

\bibitem{park2013joint}
S.-H. Park, O.~Simeone, O.~Sahin, and S.~Shamai, ``Joint precoding and multivariate backhaul compression for the downlink of cloud radio access networks,'' \emph{IEEE Transactions on Signal Processing}, vol.~61, no.~22, pp. 5646--5658, 2013.

\bibitem{lee2016multivariate}
W.~Lee, O.~Simeone, J.~Kang, and S.~Shamai, ``Multivariate fronthaul quantization for downlink {C-RAN},'' \emph{IEEE Transactions on Signal Processing}, vol.~64, no.~19, pp. 5025--5037, 2016.

\bibitem{qiao2024meta}
R.~Qiao, T.~Jiang, and W.~Yu, ``Meta-learning-based fronthaul compression for cloud radio access networks,'' \emph{IEEE Transactions on Wireless Communications (Early Access)}, 2024.

\bibitem{sanderovich2009uplink}
A.~Sanderovich, O.~Somekh, H.~V. Poor, and S.~Shamai, ``Uplink macro diversity of limited backhaul cellular network,'' \emph{IEEE Transactions on Information Theory}, vol.~55, no.~8, pp. 3457--3478, 2009.

\bibitem{khorsandmanesh2022quantization}
Y.~Khorsandmanesh, E.~Bj{\"o}rnson, and J.~Jald{\'e}n, ``Quantization-aware precoding for {MU}-{MIMO} with limited-capacity fronthaul,'' in \emph{Proc. IEEE International Conference on Acoustics, Speech and Signal Processing (ICASSP)}, Singapore, Singapore, May 2022.

\bibitem{demir2024cell}
{\"O}.~T. Demir, M.~Masoudi, E.~Bj{\"o}rnson, and C.~Cavdar, ``Cell-free massive {MIMO} in {O-RAN}: Energy-aware joint orchestration of cloud, fronthaul, and radio resources,'' \emph{IEEE Journal on Selected Areas in Communications}, vol.~42, no.~2, pp. 356 -- 372, 2024.

\bibitem{bashar2020exploiting}
M.~Bashar, A.~Akbari, K.~Cumanan, H.~Q. Ngo, A.~G. Burr, P.~Xiao, M.~Debbah, and J.~Kittler, ``Exploiting deep learning in limited-fronthaul cell-free massive {MIMO} uplink,'' \emph{IEEE Journal on Selected Areas in Communications}, vol.~38, no.~8, pp. 1678--1697, 2020.

\bibitem{arad2018precode}
N.~Arad and Y.~Noam, ``Precode and quantize channel state information sharing for cloud radio access networks,'' in \emph{Proc. IEEE International Conference on Communications (ICC)}, Kansas City, MO, USA, May 2018.

\bibitem{wiffen2021distributed}
F.~Wiffen, W.~H. Chin, and A.~Doufexi, ``Distributed dimension reduction for distributed massive {MIMO} {C-RAN} with finite fronthaul capacity,'' in \emph{Proc. IEEE 2021 55th Asilomar Conference on Signals, Systems, and Computers}, Virtual Conference, Nov. 2021.

\bibitem{liu2015graph}
J.~Liu, S.~Zhou, J.~Gong, Z.~Niu, and S.~Xu, ``Graph-based framework for flexible baseband function splitting and placement in {C-RAN},'' in \emph{Proc. IEEE International Conference on Communications (ICC)}.\hskip 1em plus 0.5em minus 0.4em\relax IEEE, 2015, pp. 1958--1963.

\bibitem{kang2015fronthaul}
J.~Kang, O.~Simeone, J.~Kang, and S.~Shamai, ``Fronthaul compression and precoding design for {C-RAN}s over ergodic fading channels,'' \emph{IEEE Transactions on Vehicular Technology}, vol.~65, no.~7, pp. 5022--5032, 2015.

\bibitem{murti2022learning}
F.~W. Murti, S.~Ali, G.~Iosifidis, and M.~Latva-Aho, ``Learning-based orchestration for dynamic functional split and resource allocation in {vRANs},'' in \emph{Proc. Joint European Conference on Networks and Communications \& {6G} Summit (EuCNC/{6G} Summit)}, Grenoble, France, June 2022.

\bibitem{bjornson2020scalable}
E.~Bj{\"o}rnson and L.~Sanguinetti, ``Scalable cell-free massive {MIMO} systems,'' \emph{IEEE Transactions on Communications}, vol.~68, no.~7, pp. 4247--4261, 2020.

\bibitem{parida2022cell}
P.~Parida and H.~S. Dhillon, ``Cell-free massive {MIMO} with finite fronthaul capacity: A stochastic geometry perspective,'' \emph{IEEE Transactions on Wireless Communications}, vol.~22, no.~3, pp. 1555--1572, 2022.

\bibitem{lorca2013lossless}
J.~Lorca and L.~Cucala, ``Lossless compression technique for the fronthaul of {LTE}/{LTE}-advanced cloud-ran architectures,'' in \emph{Proc. IEEE 14th International Symposium on a World of Wireless, Mobile and Multimedia Networks (WoWMoM)}, Madrid, Spain, June 2013.

\bibitem{simeone2022machine}
O.~Simeone, \emph{Machine Learning for Engineers}.\hskip 1em plus 0.5em minus 0.4em\relax Cambridge University Press, 2022.

\bibitem{leighton1979graph}
F.~T. Leighton, ``A graph coloring algorithm for large scheduling problems,'' \emph{Journal of Research of the National Bureau of Standards}, vol.~84, no.~6, p. 489, 1979.

\bibitem{alet2018modular}
F.~Alet, T.~Lozano-P{\'e}rez, and L.~P. Kaelbling, ``Modular meta-learning,'' in \emph{Proc. Conference on Robot Learning (CoRL)}, Zürich, Switzerland, Oct. 2018.

\bibitem{nikoloska2022modular}
I.~Nikoloska and O.~Simeone, ``Modular meta-learning for power control via random edge graph neural networks,'' \emph{IEEE Transactions on Wireless Communications}, vol.~22, no.~1, pp. 457--470, 2022.

\bibitem{griewank1993some}
A.~Griewank, ``Some bounds on the complexity of gradients, jacobians, and hessians,'' in \emph{Complexity in Numerical Optimization}.\hskip 1em plus 0.5em minus 0.4em\relax World Scientific, 1993, pp. 128--162.

\bibitem{christensen2008weighted}
S.~S. Christensen, R.~Agarwal, E.~De~Carvalho, and J.~M. Cioffi, ``Weighted sum-rate maximization using weighted {MMSE} for {MIMO}-{BC} beamforming design,'' \emph{IEEE Transactions on Wireless Communications}, vol.~7, no.~12, pp. 4792--4799, 2008.

\bibitem{feng2021weighted}
C.~Feng, W.~Shen, J.~An, and L.~Hanzo, ``Weighted sum rate maximization of the {mmWave} cell-free {MIMO} downlink relying on hybrid precoding,'' \emph{IEEE Transactions on Wireless Communications}, vol.~21, no.~4, pp. 2547--2560, 2021.

\bibitem{gholami2022survey}
A.~Gholami, S.~Kim, Z.~Dong, Z.~Yao, M.~W. Mahoney, and K.~Keutzer, ``A survey of quantization methods for efficient neural network inference,'' in \emph{Low-Power Computer Vision}.\hskip 1em plus 0.5em minus 0.4em\relax Chapman and Hall/CRC, 2022, pp. 291--326.

\bibitem{schmidt2008minimum}
D.~A. Schmidt, M.~Joham, and W.~Utschick, ``Minimum mean square error vector precoding,'' \emph{European Transactions on Telecommunications}, vol.~19, no.~3, pp. 219--231, 2008.

\bibitem{gb08}
M.~Grant and S.~Boyd, ``Graph implementations for nonsmooth convex programs,'' in \emph{Recent Advances in Learning and Control}, ser. Lecture Notes in Control and Information Sciences, V.~Blondel, S.~Boyd, and H.~Kimura, Eds.\hskip 1em plus 0.5em minus 0.4em\relax Springer-Verlag Limited, 2008, pp. 95--110, \url{http://stanford.edu/~boyd/graph_dcp.html}.

\bibitem{kingma2014adam}
D.~P. Kingma and J.~Ba, ``Adam: A method for stochastic optimization,'' in \emph{Proc. 3rd International Conference on Learning Representations (ICLR)}, Banff, AB, Canada, May 2014.

\bibitem{3gpp}
3GPP, ``Study on channel model for frequencies from 0.5 to 100 {GHz},'' \emph{TR 38.901}, Release 16.1.

\bibitem{vu2007mimo}
M.~Vu and A.~Paulraj, ``{MIMO} wireless linear precoding,'' \emph{IEEE Signal Processing Magazine}, vol.~24, no.~5, pp. 86--105, 2007.

\bibitem{venugopal2019optimal}
K.~Venugopal, N.~Gonz{\'a}lez-Prelcic, and R.~W. Heath, ``Optimal frequency-flat precoding for frequency-selective millimeter wave channels,'' \emph{IEEE Transactions on Wireless Communications}, vol.~18, no.~11, pp. 5098--5112, 2019.

\bibitem{le2003signal}
S.~Y. Le~Goff, ``Signal constellations for bit-interleaved coded modulation,'' \emph{IEEE Transactions on Information Theory}, vol.~49, no.~1, pp. 307--313, 2003.

\bibitem{duchi2011adaptive}
J.~Duchi, E.~Hazan, and Y.~Singer, ``Adaptive subgradient methods for online learning and stochastic optimization.'' \emph{Journal of Machine Learning Research}, vol.~12, no.~7, 2011.

\bibitem{shlezinger2022model}
N.~Shlezinger, Y.~C. Eldar, and S.~P. Boyd, ``Model-based deep learning: On the intersection of deep learning and optimization,'' \emph{IEEE Access}, vol.~10, pp. 115\,384--115\,398, 2022.

\bibitem{jain2024mnemosyne}
D.~Jain, K.~M. Choromanski, K.~A. Dubey, S.~Singh, V.~Sindhwani, T.~Zhang, and J.~Tan, ``Mnemosyne: Learning to train transformers with transformers,'' in \emph{Proc. Advances in Neural Information Processing Systems (NeurIPS)}, Vancouver, British Columbia, Canada, Dec. 2024.

\end{thebibliography}
\end{document}